\documentclass[10pt,journal,compsoc]{IEEEtran}

\ifCLASSOPTIONcompsoc
  \usepackage[nocompress]{cite}
\else
  \usepackage{cite}
\fi
\ifCLASSINFOpdf
\else
\fi
\usepackage{color}
\usepackage{algorithm}
\usepackage{algpseudocode}
\usepackage{graphicx}     
\usepackage{fixltx2e}
\usepackage{amsmath}
 \usepackage{graphicx}
\usepackage{caption}
\usepackage{float}
\usepackage{graphics}
\usepackage{cite}
\usepackage{multirow}
\usepackage{xcolor,colortbl}
\usepackage{varwidth}
\usepackage{hhline}

\definecolor{Gray}{gray}{0.85}
\definecolor{LightCyan}{rgb}{0.88,1,1}
\definecolor{Green}{rgb}{0.1, 0.1,0.1}

\makeatletter
\newcommand{\algmargin}{\the\ALG@thistlm}
\makeatother
\newlength{\whilewidth}
\algdef{SE}[parWHILE]{parWhile}{EndparWhile}[1]
  {\parbox[t]{\dimexpr\linewidth-\algmargin}{%
     \hangindent\whilewidth\strut\algorithmicwhile\ #1\ \algorithmicdo\strut}}{\algorithmicend\ \algorithmicwhile}%
\algnewcommand{\parState}[1]{\State%
  \parbox[t]{\dimexpr\linewidth-\algmargin}{\strut #1\strut}}

\newcommand{\abs}[1]{\lvert#1\rvert}
\newcommand{\norm}[1]{\lVert#1\rVert}
\begin{document}
%

\title{Application Level High Speed Transfer Optimization Based on Historical Analysis and Real-time Tuning}

%
%
%
%

\author{Engin~Arslan~and~Tevfik~Kosar
\IEEEcompsocitemizethanks{\IEEEcompsocthanksitem E. Arslan is with University of Illinois at Urbana Champaign.\protect\\
E-mail: engin@illinois.edu
\IEEEcompsocthanksitem T. Kosar is with University at Buffalo, SUNY.}
\thanks{}}

\newcommand{\algoName}{HARP}
\IEEEtitleabstractindextext{%
\begin{abstract}

Data-intensive scientific and commercial applications increasingly require frequent movement of large datasets from one site to the other(s). Despite growing network capacities, these data movements rarely achieve the promised data transfer rates of the underlying physical network due to poorly tuned data transfer protocols. Accurately and efficiently tuning the data transfer protocol parameters in a dynamically changing network environment is a major challenge and remains as an open research problem. In this paper, we present predictive end-to-end data transfer optimization algorithms based on historical data analysis and real-time background traffic probing, dubbed {\algoName}. Most of the previous work in this area are solely based on real time network probing which results either in an excessive sampling overhead or fails to accurately predict the optimal transfer parameters. Combining historical data analysis with real time sampling enables our algorithms to tune the application level data transfer parameters accurately and efficiently to achieve close-to-optimal end-to-end data transfer throughput with very low overhead. Our experimental analysis over a variety of network settings shows that {\algoName} outperforms existing solutions by up to 50\% in terms of the achieved throughput.

\end{abstract}

}

\maketitle

\IEEEdisplaynontitleabstractindextext

%
\IEEEpeerreviewmaketitle

\ifCLASSOPTIONcompsoc
\IEEEraisesectionheading{\section{Introduction}\label{sec:introduction}}
\else
\section{Introduction}
\label{sec:introduction}
\fi
As the trend towards more data-intensive applications continues, developers and users need to invest significant effort 
into efficiently moving large datasets between distributed sites. 
Effective use of the available network bandwidth together with optimization of data transfer throughput 
have been critical for the end-to-end performance observed by most commercial and scientific applications. This is true 
despite multi-gigabit optical network offerings. Most users fail to 
obtain even a fraction of the theoretical speeds promised by existing
networks due to issues such as sub-optimal end-system and network protocol tuning. 


Most of the existing work on data transfer tuning and optimization is at the low-level, including design of new transport protocols~\cite{Eggert:2000:EE:505688.505691, tcprome, Crowcroft:1998:DEI:293927.293930, Kola:2007:TBS:1284907.1285024} as well as adapting and changing the existing transport protocols for better performance~\cite{R_Lee01, R_Kola05}. At a higher level, other techniques have been developed by keeping the existing underlying protocol intact and tuning it at the application level for improved performance.
One common way to address protocol tuning at the application level is through the tuning of parameters such as pipelining~\cite{TCP_Pipeline, farkas2002}, parallelism~\cite{R_Lee01, R_Hacker05, R_Karrer06, R_Lu05}, concurrency~\cite{kosar04, Kosar09, R_Liu10}, and buffer size~\cite{R_Prasad04, R_Morajko04, R_Ito08, R_Choi05}. 
These parameters can be tuned at the application level without the need for changing the underlying transfer protocols and can significantly improve the end-to-end data transfer performance~\cite{europar13,Esma-DADC09}.



While significant performance gain can be achieved by tuning application level protocol parameters, the optimal value of these transfer parameters varies depending on the dataset (i.e., file size and the number of files), network (i.e., bandwidth, round-trip-time, and background traffic on network), and end-system characteristics (i.e., file system and transfer protocol chosen). Thus, finding the best combination for these parameters is a challenging task. 
For instance, pipelining helps in transferring multiple files back-to-back without waiting for an acknowledgement message in the control channel, but the size of the transferred files must be small to benefit from this approach. Pipelining may even cause the throughput to decrease when set to high values for large files. Instead, large files would benefit from being divided into smaller chunks and being transferred over the network through multiple parallel streams. Similarly, both small and large files would benefit from concurrency, meaning simultaneously transferring multiple files over different transfer streams/channels. On the other hand, opening too many connections to transfer a single file (parallelism) or multiple files (concurrency) would also degrade the throughput by increasing network congestion and causing oversubscription in the file system. Optimal values for these parameters  depend on the many factors described above. In our previous work, we have developed heuristic-based dynamic optimization algorithms~\cite{europar13} to determine  the best combination of these parameters by using network and dataset characteristics (i.e., bandwidth, round-trip-time, and average file size).


\IEEEpubidadjcol

In this paper, we present predictive end-to-end data transfer optimization algorithms based on historical data analysis and real-time background traffic probing (\algoName). 
We use historical data to derive network specific models of transfer throughput based on protocol parameters. Then by running sample transfers, we capture the current load on the network which is fed into these models to increase the accuracy of our predictive modeling. Combining historical data analysis with real time sampling enables our algorithms to tune the application level data transfer parameters (i.e., parallelism, pipelining, and concurrency) accurately and efficiently to achieve close-to-optimal end-to-end data transfer throughput with very low sampling overhead. Our experimental analysis over a variety of network settings shows that {\algoName}~outperforms existing solutions by up to 50\% in terms of achieved throughput. We also propose ``online tuning'' that monitors transfer throughput periodically and updates values of protocol parameters when network conditions change such as variability in background traffic. Online tuning is able to increase the data transfer throughput by up to 30-40\% compared to~\algoName~without online tuning.

The rest of this paper is organized as follows: Section II motivates our work; Section III presents our system design and the proposed algorithms; Section IV discusses the evaluation of our model; Section V describes the related work in this field; and Section VI concludes the paper with a discussion on the future work.


%
%
%
%

\section{Motivation} \label{sec:motivation}
Tunable transfer parameters such as pipelining, parallelism, and concurrency play a significant role in improving the achievable transfer throughput.
However, setting the optimal levels for these parameters is a challenging task and an open research problem. Poorly-tuned parameters can either cause underutilization of the available network bandwidth or overburden the network links and degrade the performance due to increased packet loss, end-system overhead, and other factors.

Among these parameters, {\em pipelining} targets the problem of transferring a large numbers of small files~\cite{TCP_Pipeline, farkas2002}.
In most control channel-based transfer protocols, an entire transfer must complete and be acknowledged before the next transfer command is sent by the client.
This may cause a delay of more than one round-trip-time (RTT) between the individual transfers.
With pipelining, multiple transfer commands can be queued  at the server, greatly reducing the delay between transfer completion and the receipt of the next command.
{\em Parallelism} sends different chunks of the same file over parallel data streams (typically TCP connections),
and can achieve high throughput by aggregating multiple streams and utilizing a larger share of the available network bandwidth~\cite{R_Hacker05, R_Lu05, DADC_2008, NDM_2011}.
{\em Concurrency} refers to sending multiple files simultaneously through the network using different data channels, and is especially useful for increasing I/O concurrency in parallel disk systems~\cite{Thesis_2005, NDM_2012, JGrid_2012}.

\begin{figure}[t]
\begin{center}
\includegraphics[keepaspectratio=true,angle=0,width=72mm] {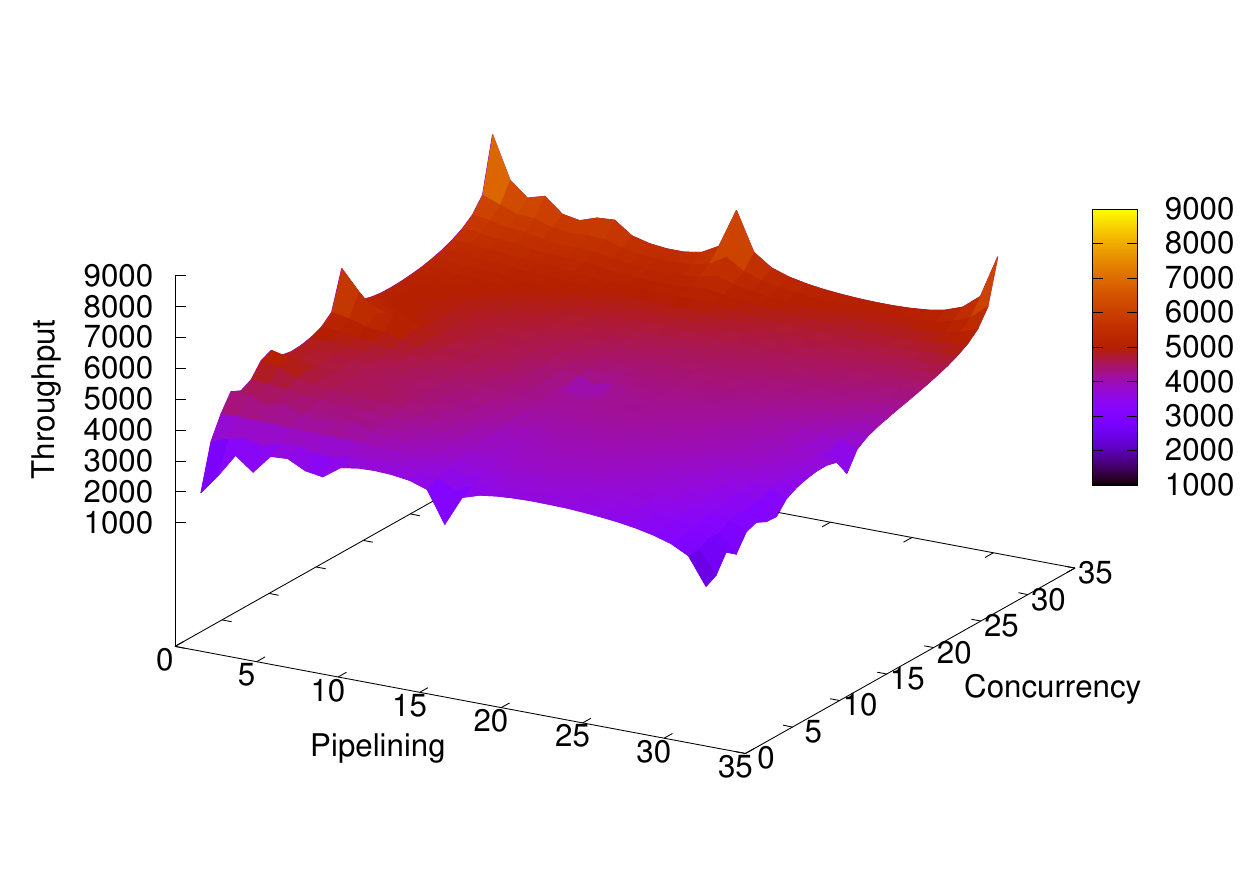}
\vspace{-2mm}
\caption{Large pipelining level when combined with concurrency leads up to 20\% less throughput.}
\label{fig:param-effect}
\end{center}
\vspace{-6mm}
\end{figure}

Figure~\ref{fig:param-effect} shows the impact of protocol parameters concurrency and pipelining on transfer throughput. When small concurrency values are used, larger pipelining values achieve better throughput than default value 1. On the other hand, as concurency value is increased, pipelining starts impacting negatively. When concurrency is set to 32 and pipelining is set to 1 the transfer throughput becomes 8120 Mbps. When pipelining is increased to 8, 16 and 32 for the same concurrency value, throughput becomes 6730, 7300, 7290 respectively. The reason pipelining starts causing negative impact when concurrent is large is that it causes disproportionate allocation of files to channels. When there are more than one channels available (which refers more than one concurrency level), then assigning files to channels in advance may cause some channels to finish its assigned files in case its files are larger than others.

Furthermore, background traffic may change while transfer is running especially if transfer lasts long such as hours or days. Since optimal parameter values for the new background traffic is different than what has been set at the beginning of the transfer, updating parameter values would be necessary to achieve sustained high transfer throughput. In Section~\ref{sec:onlinetuning}, we propose ``online tuning'' to monitor transfer throughput periodically and update parameter values if network condition has changed under which circumstance the optimal parameter values would be different than currently used ones.

\section{Overview of \algoName} \label{sec:design}
\algoName~combines three approaches of application-level data transfer tuning and optimization: {\em i)} heuristics; {\em (ii)} real-time probing; and {\em (iii)} historical data analysis. Heuristic algorithms~\cite{globusonline, europar13} compute transfer parameters through calculations on the dataset and network metrics. For example, the value of pipelining is calculated by dividing the bandwidth-delay-product (BDP) to the average file size so that it will return large values for small files and small values for large files which aligns with the purpose of pipelining~\cite{R_Bres07}. However, heuristics fail to capture the dynamic changes in the network and end-system specific settings, including the real-time background traffic. Real-time probing based approaches~\cite{esma-tcc, Esma-DADC09} find optimal values of protocol parameters by running a sequence of sample transfers (probes) using different metric values. They take a portion of the original dataset and transfer it with initial metric value (generally 1), followed by a series of sample transfers with increased values (2, 4, 8, etc.) until the observed throughput stops increasing. Although real-time probing has an advantage of capturing the instantaneous network load, it may bring too much probing overhead to accurately discover the optimal values of parameters. Finally, historical data methods~\cite{zulkar-ndm14, raj-ccgrid14} model data transfer throughput based on dataset, network, and protocol metrics. In order to capture change in the network load, they either rely on recent historical data~\cite{raj-ccgrid14} or run sample transfers~\cite{zulkar-ndm14}. \algoName~runs sample transfers similar to probing based solutions, but the number of sample transfers are way less than it is in probing based algorithms. \algoName~benefits from heuristic solutions to determine parameter values of sample transfers. Since poor choice of parameter values in sample transfer may affect overall throughput, taking advantage of heuristic algorithms may alleviate the sample overhead. Finally, \algoName~models transfer throughput similar to historical data based solutions. While models driven by Kettimuthu et al.~\cite{raj-ccgrid14} and Nine et al.~\cite{zulkar-ndm14} require offline analysis and can work well only for networks for which they are trained, \algoName~requires no prior data analysis and can be applied to different networks with the help of extensible similarity detection algorithm.


\begin{figure}[t]
\begin{centering}
 \includegraphics[keepaspectratio=true,angle=0,width=80mm]{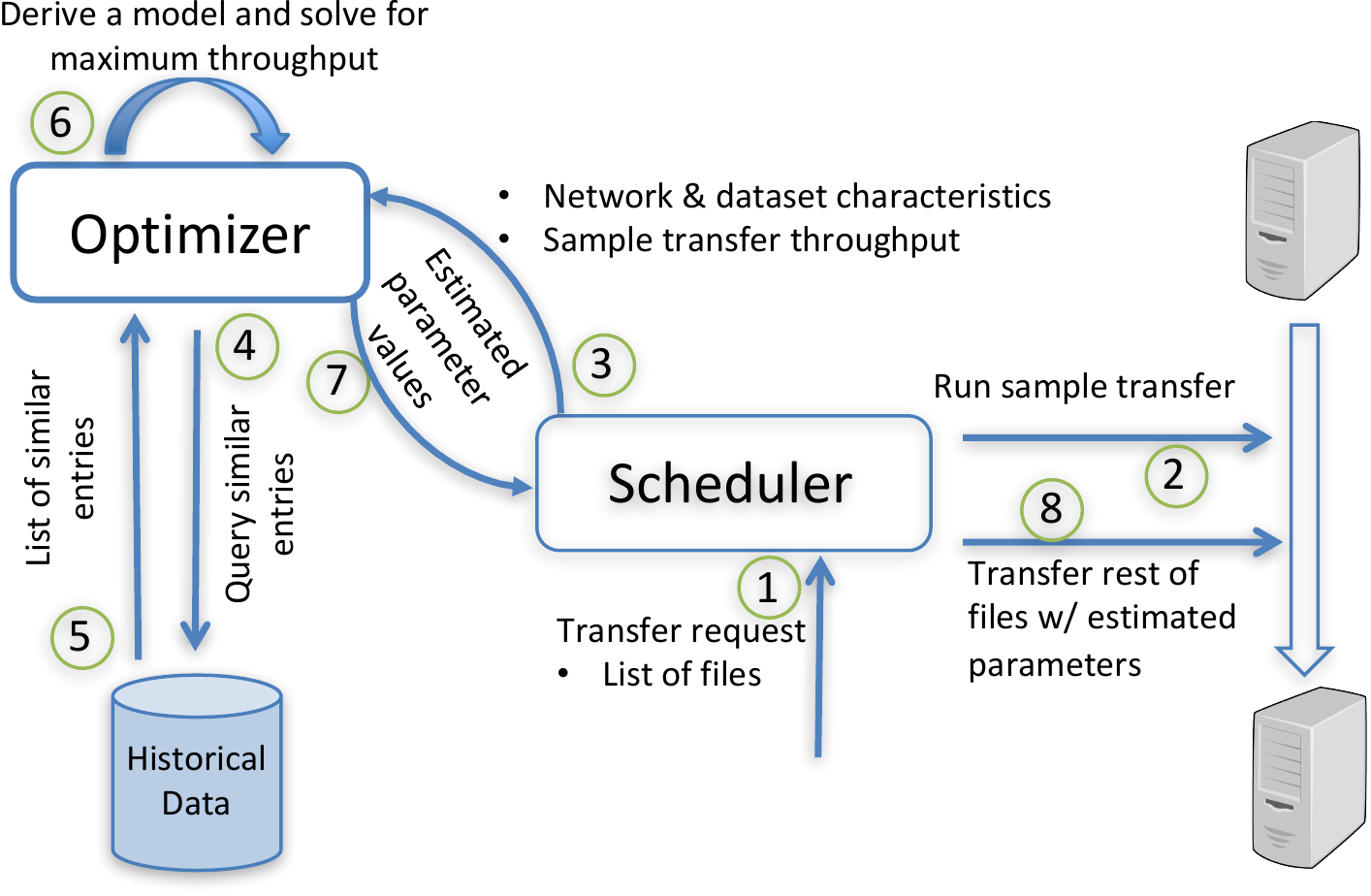}
\caption{Flow of operations in \algoName.} \label{fig:architecture}
\end{centering}
\vspace{-3mm}
\end{figure}

\algoName~is composed of two main modules which are Scheduler and Optimizer as shown in Figure~\ref{fig:architecture}. When a data transfer request is submitted to the Scheduler, it first categorizes files in the transfer request into groups based on the file size. Then, it runs one sample transfer for each file group to capture the load on the network (step 2) as well as the effect of the network load on the file groups. Once the sample transfer throughputs are obtained, it passes this information along with the network/dataset characteristics (step 3) to  Optimizer to determine the similar entries in the historical data (step 4 and 5). Then, Optimizer identifies similar entries and runs regression analysis to derive a model that relates transfer parameters to the transfer throughput. The derived model is then solved for the maximum transfer throughput and corresponding parameter values are obtained (step 6). After parameter values are found, parameter relaxation process is used to lower the values of parameters while keeping the estimated throughput in a reasonable range. Finally, it returns the parameter combinations to Scheduler (step 7) to schedule the transfer of the rest of the dataset with the calculated parameter values (step 8).

\begin{algorithm}[t]
\scriptsize
\centering
\caption{-- Scheduler of \algoName}
\begin{algorithmic}[1]
\Statex
\Function{transfer}{source,destination,BW,RTT, algorithm}
	\State $allFiles = getListOfFiles()$
	\State $chunks = partitionFiles(allFiles)$\label{line:partition}
	\For{$i=0$; $i<chunks.length$; $i++$}
		\State $sampleFiles = chunks[i].split(SAMPLING\_SIZE)$
		\parState {$(cc,p, pp) = findParamsViaHeuristic(sampleFiles, BW, RTT)$}
		\parState {$ST[i] = transferChunk(sampleFiles, cc,p, pp)$}\label{line:sampling}
	\EndFor
	\State $maxCC = 1$
	\State $TT = 0$
	\For{$i=0$; $i<chunks.length$; $i++$}
		\State \begin{varwidth}[t]{\linewidth}
		$cc_{est}[i], p_{est}[i], pp_{est}[i], UT[i] = runOptimizer(ST[i],$\par
		\hskip\algorithmicindent $BW, RTT, metadata(chunk[i]))$\label{line:runOptimizer}
		\end{varwidth}
		\State $TT \mathrel{+}= UT[i]$
		\State $maxCC = max(maxCC, cc_{est}[i])$\label{line:conc-max}
	\EndFor
	\For{$i=0$; $i<chunks.length$; $i++$}
			\parState { $weight[i] = chunks[i].size * \frac{TT}{UT[i]}$}\label{line:promc-weight}
			\State $totalWeight \mathrel{+}= weight[i]$
	\EndFor
	\For{$i=0$; $i<chunks.length$; $i++$}
			\State $cc' = max(cc_{est}[i], \left\lfloor maxCC * \frac{weight[i]}{totalWeight} \right\rfloor)$ 
			\State $startTransfer(chunks[i], cc', p_{est}[i], pp_{est}[i])$ \Comment Asynchronous operation
	\EndFor
\EndFunction
\end{algorithmic}
\label{alg:scheduler}
\end{algorithm}

\subsection{Transfer Scheduler}\label{sec:scheduler}
\algoName's~Scheduler is responsible for managing data transfer executions between end points. It first divides files into groups (a.k.a., chunks) according to the file size (i.e., Tiny, Small, Medium, and Large) (line~\ref{line:partition} of Algorithm~\ref{alg:scheduler}). Then, it runs one sample transfer (a.k.a., probing) for each chunk to learn about achievable throughput of each chunk. $ST[i]$ refers to sample transfer throughput for $chunk_i$. In order to minimize the overhead of sample transfers, Scheduler takes advantage of heuristics~\cite{europar13} to determine the parameter values of the sample transfer. Although heuristic calculations may pick suboptimal values, it is generally better than default or random values. We have explained the cost of sample transfer in Section~\ref{sec:cost} in detail. 

After real-time probing is completed, Scheduler sends the achieved throughput to Optimizer along with the dataset and network settings (line~\ref{line:runOptimizer}). Optimizer returns values for protocol parameters $(cc_{est}, p_{est}, ppq_{est})$ along with unit throughput, $UT$ (line 12). $UT$ refers to throughput of a chunk if it is run with concurrency value 1. It is used to determine how channels will be distributed among chunks when chunks are run concurrently. While Optimizer finds optimal values assuming each chunk will run separately, Scheduler prefers to run multiple chunks simultaneously to benefit from multi-chunk approach as presented in our earlier work~\cite{europar13}. While we can use parallelism and pipelining values as returned by Optimizer, concurrency must be adapted to multi-chunk transfer scheme. Even though concurrency has significant impact on throughput (especially when parallel file systems are in use), it also causes the highest overhead at the end systems and network by creating multiple processes. Thus, it may not be possible, yet undesirable, to open as many channels (concurrency) as each chunk asks. To overcome this inconsistency, Scheduler computes the maximum concurrency ($maxCC$) of all chunks (line~\ref{line:conc-max}) which is then distributed among chunks based on their weights (line 21). Weight of a chunk is proportional to size of a chunk and inversely proportional to ratio to $\frac{UT}{TT}$ where $TT$ refers to sum of all $UT$s (line 17). 

To give an example of channel distribution, let's assume we have three chunks (aka. file types) in a given dataset and each chunk's total size is 1 GB. Let's also say, Optimizer returned (7,1,10,100), (4,3,1,200), and (3,5,0,400) as (cc,p,pp,UT) combination. Then, $TT$ will be 700 and $maxCC$ will be 7 (maximum of (7,4,3)). Based on weight calculation method shown in line~\ref{line:conc-max}, 7x, 3.5x, and 1.75x weights will be assigned to chunks, respectively. Finally, when maximum concurrency value is distributed to chunks based on respective weights, first chunk will receive four concurrency, second chunk will receive two concurrency and last chunk will receive one concurrency. Since parallelism and pipelining values returned by Optimizer are used as is, final protocol parameter values for chunks will be (4,1,10), (2,3,1), and (1,3,0). Once final parameter values of chunks are determined, Scheduler runs them concurrently.

\subsubsection{Adaptive Sample Transfers}\label{sec:sample_transfer}

\begin{figure}[t]
\begin{center}
\includegraphics[keepaspectratio=true,angle=0,width=72mm] {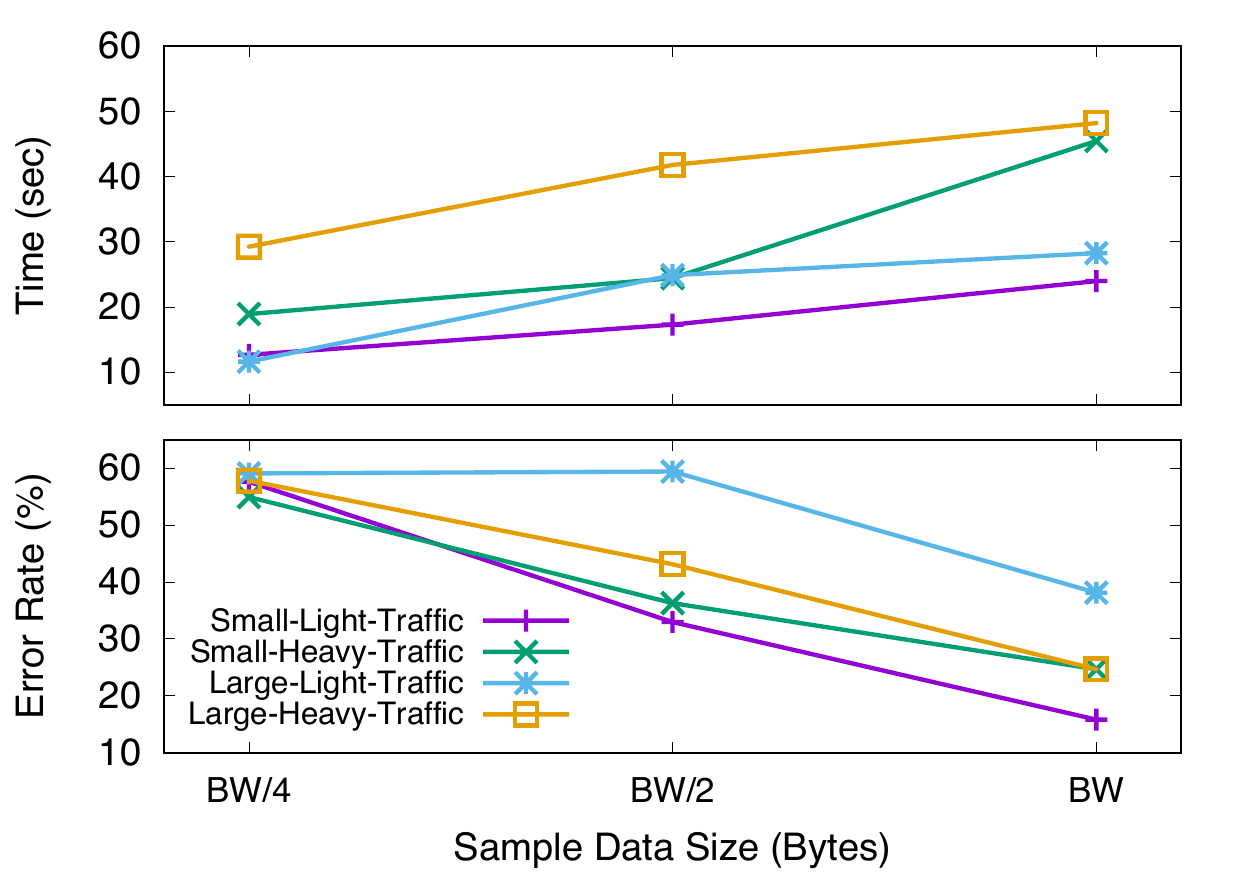}
\caption{Sampling with fixed data size causes long delay and high error rates.}
\label{fig_nonadaptive}
\end{center}
\vspace{-4mm}
\end{figure}

Two approaches have been proposed so far for running sample transfers. They are (i) fixed data size~\cite{esma-tcc} and (ii) fixed time duration~\cite{power-aware} based sampling. In the first method, a fixed-size (e.g. 1GB) portion of a original dataset is used to run sample transfers, however finding optimal data size for sample transfer requires prior analysis~\cite{esma-sampling}. Thus, we have compared three data sizes under two background traffic condition and measured accuracy and time spent in Figure~\ref{fig_nonadaptive}. Instead of using static values, we selected data sizes proportional to network bandwidth. For example, {\it BW} data size will use 10 GB of original dataset if network bandwidth is 10 Gbps such that in an ideal condition the sample transfer will finish in 8 seconds. To measure accuracy, we compared throughput of sample transfers to throughput when sample size is set to total dataset size.

Error rate of sample transfer decreases from ~50-60\% to 20-40\% as sample size increases from $BW/4$ to $BW$ as shown in Figure~\ref{fig_nonadaptive}. Larger sample sizes, on the other hand, cause longer sample transfer duration as expected. Since goal of a sample transfer is to estimate achievable transfer throughput with minimal overhead, using too large sample sizes could cause overall performance degradation in cases sample transfer parameters are far from optimal hence it takes too long to finish. Heavy background traffic exacerbates the situation and increases sample transfer time up to 50 seconds. Thus, using fixed size data to run sample transfer is not an optimal method as small size causes high error rates and large size induces high transfer times. A key reason why error rate of sample transfer is high in fixed-size approach is that sample transfer spends considerable amount of time during user authentication, connection establishment/tear-down and slow start phase which causes underestimation of actual throughput. On the other hand, fixed-time approach requires fine tuning of the time duration during which the sample transfer will run. That is, using short time duration for long RTT networks would cause higher error rates and large time duration would lead longer delay in sampling phase. Yet, similar to fixed-size method, the error rate of fixed-time approach will also be affected by transfer startup/tear-down operations. Hence, we propose a novel adaptive sample transfer method which monitors sample transfer throughput periodically and stops transfer as soon as it observes convergence in transfer throughput. 

\begin{figure*}[t]
\centering
\includegraphics[keepaspectratio=true,angle=0,width=56mm] {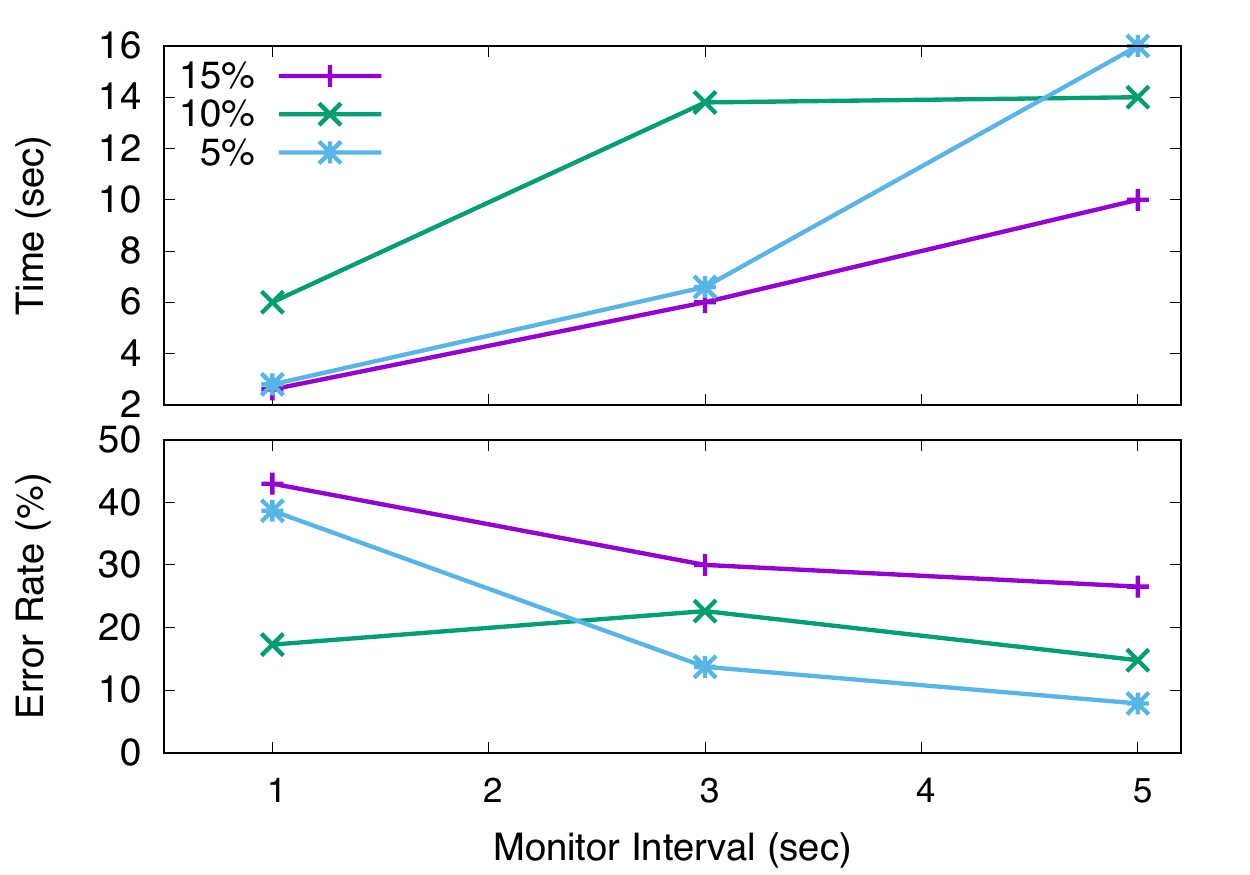}
\includegraphics[keepaspectratio=true,angle=0,width=56mm] {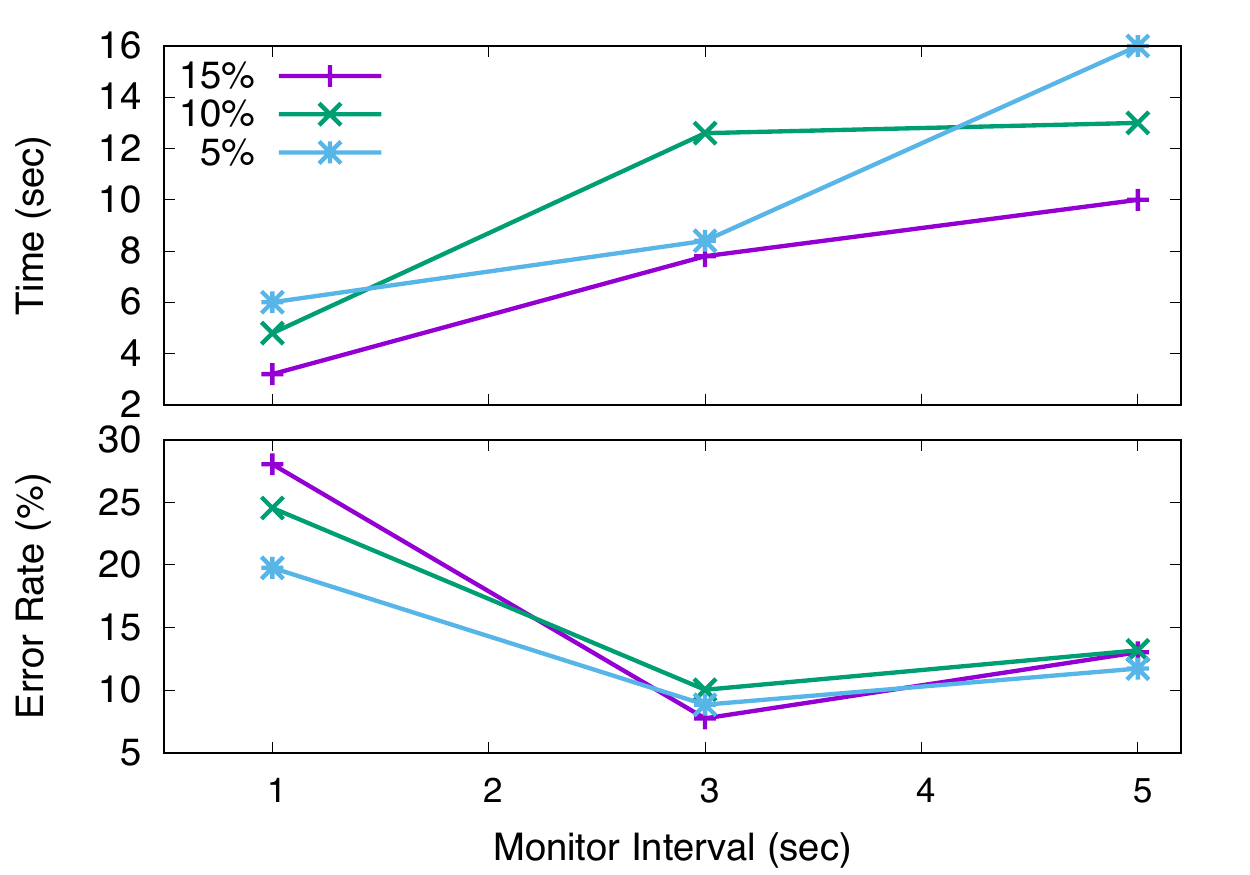}
\includegraphics[keepaspectratio=true,angle=0,width=56mm] {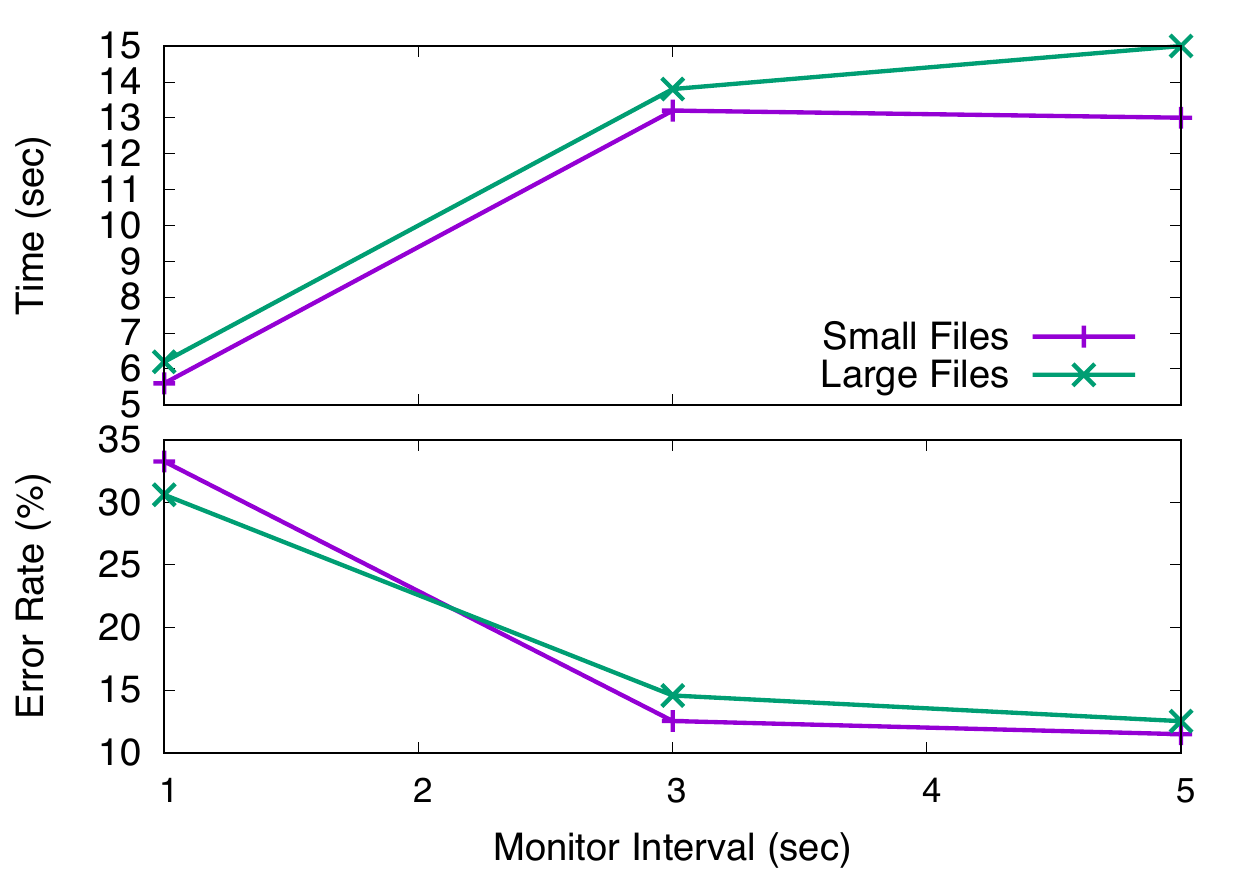}
\caption{Adaptive sample transfers can achieve higher accuracy within shorter amount of time.}
\label{fig:xsede}
\end{figure*}

In our adaptive sample transfer method, Scheduler starts transferring entire dataset and monitors instantaneous transfer throughput at certain intervals. If throughput of two consecutive monitor intervals are closer than threshold, then Scheduler exits sample transfer phase and takes the average of two consecutive intervals' throughput as throughput of the sample transfer. Rather than using static values for threshold (which needs to be adapted to different network bandwidths), we define threshold in terms of percentage of last monitor interval. That is, if throughput of current interval is $x\%$ closer to previous interval's throughput, then convergence condition is assumed to be satisfied. We have tested three values for $x$ in  Figure~\ref{fig:xsede}(a) and Figure~\ref{fig:xsede}(b) for Small and Large file types for different monitor intervals (1, 3, 5 seconds). As the threshold percentage increases, the sample transfer time shortens in exchange for higher error rates for both file types. 5\% threshold with 3 seconds monitor interval is able to achieve ~10\% error rate with less than 10 seconds delay. Compared to fixed-size sampling approach, it is 2-4X faster and more accurate. In Figure~\ref{fig:xsede}(c), we have tested 5\% threshold under heavy background traffic. Although sample time and error rates increase a little bit, it is able to achieve less than 15\% error rate in 15 seconds which is almost twice better than fixed-size method. We validated that 5\% threshold with 3 seconds monitor interval returns similar results in local area experiments but omitted them due to space limitation.

\begin{table}
\begin{centering}
\begin{tabular}{ |c| c| c| c|}
\hline
 \multirow{2}{*}{\bf Specs} &\multicolumn{1} {|c|} {XSEDE} & \multicolumn{1}{|c|} {DIDCLAB} & \multirow{2}{*} {EC2}\\
& Stampede-Gordon & WS1-WS-2 & \\
\hline
{\bf Bandwidth (Gbps)} &  10 & 1  & 10 \\
\hline
{\bf RTT (ms)} & 40 & 0.2 & 100  \\
\hline
{\bf TCP Buffer Size (MB)}& 32 & 4 & 60  \\
\hline
{\bf BDP (MB)} & 48 & 0.02 & 125 \\
\hline
{\bf File System} & Lustre & NFS  & SAN\\
\hline
{\bf Max File System } &  \multirow{2}{*} {1200} &  \multirow{2}{*}{90} &  \multirow{2}{*}{320} \\
{\bf Throughput  (MB)} & & & \\
\hline
\end{tabular}
\caption{Network specifications of the test environment.} \label{tab:system-spec}
\end{centering}
\vspace{-6mm}
\end{table}

\subsection{Optimizer} \label{sec:optimizer}
Optimization module of \algoName~aims to determine the optimal parameter values for the transfer metrics for an intended data transfer with the help of historical data. Thus, it heavily depends on the quality and quantity of the dataset to make the best decisions. Quality stands for how well the dataset captures variation in the network such as the background traffic on the network. Quantity is important (i) to detect outliers and (ii) to derive accurate models.

\subsubsection{Data Collection}
We collected our historical transfer data on XSEDE~\cite{XSEDE}, a production-level high-speed WAN, and DIDCLAB at UB, a dedicated LAN. The network and storage configurations of the used systems are given in Table~\ref{tab:system-spec}. Four different file sizes are used which are Tiny (varying from 1MB to 5MB with a total of 10GB), Small (15MB--30MB with a total of 20GB), Medium (50MB--200MB with a total of 40GB), and Large (1GB--5GB with a total of 98 GB). 

In order to observe the effect of protocol metrics under different network loads, we tested same parameter combinations under different network loads; light, medium, and heavy background traffic. Although we can control background traffic in LAN experiments, we cannot know the exact background traffic in a shared production networks, such as XSEDE. However, we observed higher and more stable throughput values when we run transfers in the night hours.  Thus, data entries are collected in the night hours of the day to minimize the effect of external load. Moreover, we repeated each entry at least five times at different dates so that any outliers can be detected and neglected easily.
For the medium and heavy background traffic cases, we synthetically created background traffic by running multiple memory-to-memory transfers in the background during the data collection. 
Over a 12-week period, we collected statistics for 21K data transfers. Since we only kept metadata (i.e., number of files, average file size, date etc.) of data transfers, the amount of storage to store historical data was around 4MB. Even though we did not experience storage limitation in our experiments, one can put a time limit to remove old entries and keep historical data size at a certain level. 

\begin{figure*}[t]
\includegraphics[width=\textwidth,height=\textheight,keepaspectratio]{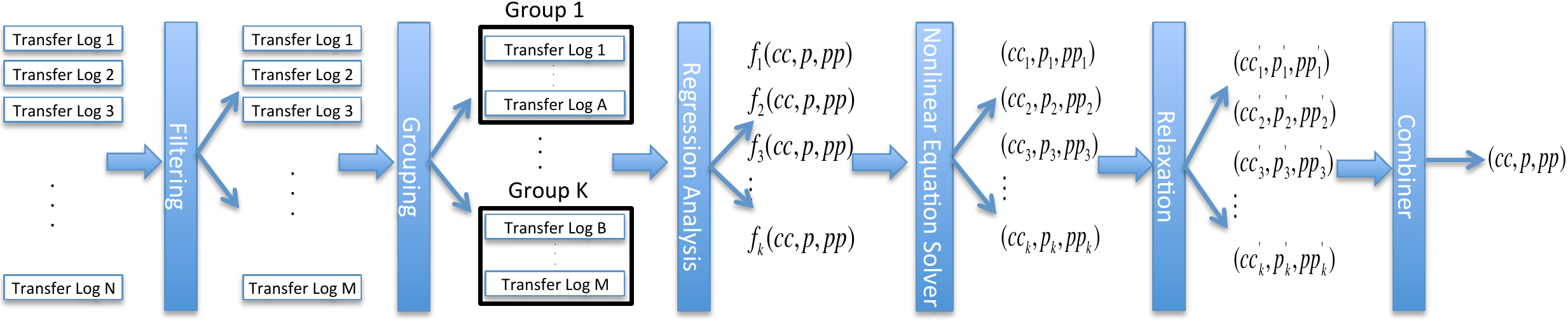}
\caption{Flow of operations in \algoName's Optimizer.} 
\label{fig:optimization-module}
\vspace{-2mm}
\end{figure*}

\subsubsection{Data Filtering and Grouping}
When Optimizer receives a request from Scheduler which consists of a dataset, network characteristics and sample transfer throughputs, the first operation it does is to filter similar entries from the data store where historical data is kept. Since it is possible that the data store may not have exactly matching entries for a given dataset/network characteristics, Optimizer uses a weighted cosine-similarity function (shown in Equation~\ref{cosine-similarity}) to measure the similarity of historical data entries to the intended transfer.
\begin{equation}
cos(\theta) = \frac{\sum\limits_{i=1}^n \norm{A_i} \norm{B_i}}{\sqrt{\sum\limits_{i=1}^n \norm{A_i^2}}\sqrt{\sum\limits_{i=1}^n \norm{B_i^2}}}
\label{cosine-similarity}
\end{equation}
Cosine-similarity uses set of features and calculates the degree of alikeness based of how the features of objects are close to each other. In Equation~\ref{cosine-similarity}, $A$ and $B$ refers to values for feature set of two instances. In the context of data transfer, feature set consists of dataset and network settings of transfers {e.g., bandwidth, round-trip-time, $\frac{bandwidth-delay-product}{buffer-size}$, chunk type (Tiny, Small, etc.), file size, and file count}. $\frac{bandwidth-delay-product}{buffer size}$ is used to determine if use of parallel streams would help to overcome TCP buffer size limitation. Even though some of the features are related to each other, such as the chunk type and the file size, we wanted to be as much specific as possible in terms of similarity detection. For example, if BDP is 40 MB, 1MB and 1KB sized files will be put into Tiny chunk and 1GB will be in Large chunk according to our dataset partitioning method. However, if we just use file size to compare similarities, 1MB file size will have same similarity value when compared to 1KB and 1GB files. To address such misclassification, we evaluate some features in multiple ways to have more accurate similarity detection. Moreover, we normalize feature vectors of historical data entries to try to keep ranges of properties as close as possible otherwise, one property may overweight the similarity value if value of a property is larger than others. Finally, since each feature has different impact on file transfer throughput, we assigned weights to them based on our initial effort to apply regression to the entire historical data. Although the accuracy of the regression is low, it gives a clue about the weight of each property on achieved transfer throughput. Hence, we used (2,2,10,10,3,1) weight values for the feature set {Bandwidth, RTT, $\frac{BDP}{buffer size}$, chunk type, file size and file count), respectively. Since there are other factors aside from dataset and network characteristics that affect the transfer throughput (e.g., background traffic, disk I/O performance, etc.), Optimizer runs additional step of categorization using the results of sample transfers which is explained in Section~\ref{sec:combiner}.

Once Optimizer calculates the similarity value (according to the intended transfer) for each entry in the historical data, it picks entries with similarity value larger than the threshold. We initialized the threshold value to 0.99 and decreased it until we have at least 6K entries. Since cosine-similarity does not consider background traffic, the selected entries will have transfers with different background traffic. To overcome the hidden variable problem, we grouped set of entries that are not only same in regard to dataset and network characteristics but also collected at approximate times. 
Although it is possible that while some of the entries that are collected similar time period are exposed to different background traffic due to transient traffic events, those will be identified and ignored in the modeling phase. During data collection, we ran each dataset with all possible combination of parameter values (from (1,1,1) to (32,32,32)). Thus, when we group historical data entries based on data collection time, each group will have 216 entries. Thus, Optimizer will have at least 30 groups (set of entries) (6K/ 216) at the end of the similarity detection phase.

\begin{figure} 
\begin{centering}
\includegraphics[keepaspectratio=true,angle=0,width=72mm]{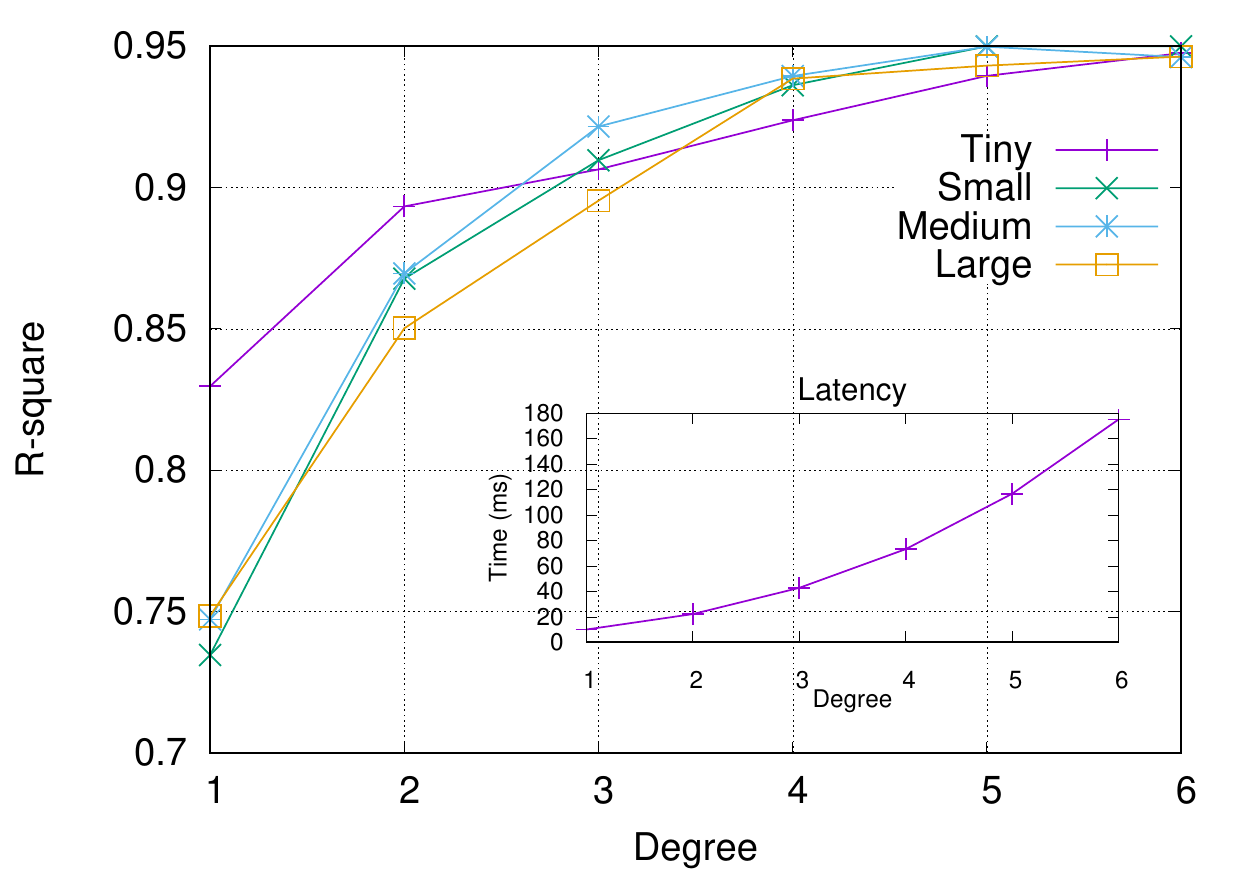}
\caption{Higher degree polynomial regressions return higher accuracy in return for longer computation time.} \label{fig:regression}
\end{centering}
\vspace{-2mm}
\end{figure}

\subsubsection{Regression Analysis and Nonlinear Equation Solver}
After similar entries are grouped based on time information, entries in same group share the same network and dataset characteristics but differ in values of protocol parameters. Hence, we can ignore dataset and network characteristics and derive a model on this data that relates protocol parameters to transfer throughput as shown in Equation~\ref{eq:poly-eq}. $Thr_i$ refers to the equation derived for $i^th$ historical data group where $1< i < N$ and $N$ is the total number of historical data groups after filtering phase. By grouping entries with the same network/dataset metrics and similar background traffic, we can decrease the number of input parameters to the model which leads to higher success in fitting regression. In order to validate regression quality, we divide each group into two subgroups as training (70\%) and validation (30\%) sets. 

We compared several degrees of polynomial regression in terms of accuracy and speed in Figure~\ref{fig:regression}. According to the results, as we increase the degree of polynomial regression, the $R^2$ value increases as well as the time to compute the model. So, we start with degree one to compute regression and increase the degree until $R^2$ for training and validation data goes above 0.7. If $R^2$ is still not larger than 0.7 for training and validation data of a group, then we classify the group as outlier and ignore it. We have observed that quadratic or cubic regressions satisfy the requirement for most of the groups to pass the outlier detection test.

\vspace{-3mm}
\begin{equation}
T_i = f_i(cc, p, pp)
\label{eq:poly-eq}
\vspace{-2mm}
\end{equation}

After polynomial equations are derived for each group of historical data, {$f_1,f_2,..., f_k$}, Optimizer evaluates them for the values used in the sample transfers to find the estimated throughputs, {$T_1$,$T_2$,...,$T_k$} as shown in Equation~\ref{eq:evaluate}. $\epsilon_i$ refers to the difference between throughputs estimated by $f_i$ and obtained by the actual sample transfer, $Thr_{act}$. In order to prioritize historical data entries that are exposed to similar background traffic compared to current traffic, Optimizer assigns weights to the equations based on their accuracy in estimating throughput of sample transfers. In order to assign weights to equations, we classify them into groups using density based clustering technique, DBScan based on $\epsilon$ values. Each equation in a class is assigned the same weight ($w$) and the weight of a class is calculated as ${2^0,2^1,...,2^{k-1}}$ where classes are sorted in descending order based on $\epsilon$ values. Weights play significant role in (i) distinguishing historical data collected under different background traffics and (ii) compensating similarity based filtering for possible misclassification due to unacknowledged factors of data transfers such as background traffic. We have observed that DBScan generally returns 4-6 groups.
\vspace{-1mm}
\begin{equation}
\epsilon_i = T_{act} - f_i(cc_0, p_0, pp_0)
\label{eq:evaluate}
\vspace{-1mm}
\end{equation}

After the weights of equations are found, Optimizer finds parameter combination for each equation that returns the highest throughput using the non-linear optimization solver with Broyden - Fletcher - Goldfarb - Shanno (BFGS) algorithm. It scans bounded solution space for the variables, $(cc_i, p_i, pp_i)$ for maximum throughput, $Tmax_i$, as shown in Equation~\ref{eq:maximize}. Optimizer combines the values found by each equation by taking weighted average as shown in Equation~\ref{eq:average}. 

\begin{figure*}[t]
\centering
\includegraphics[keepaspectratio=true,angle=0,width=60mm] {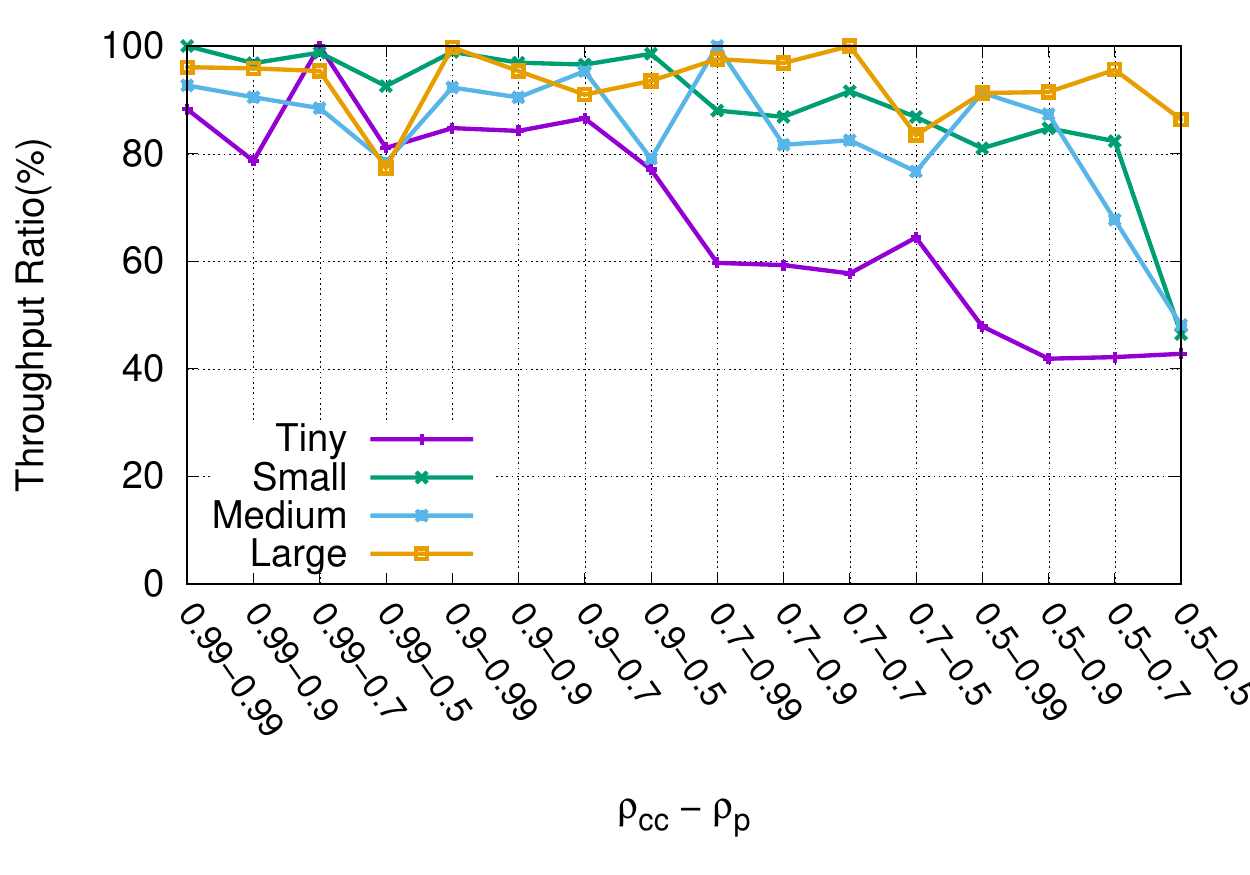}
\includegraphics[keepaspectratio=true,angle=0,width=60mm] {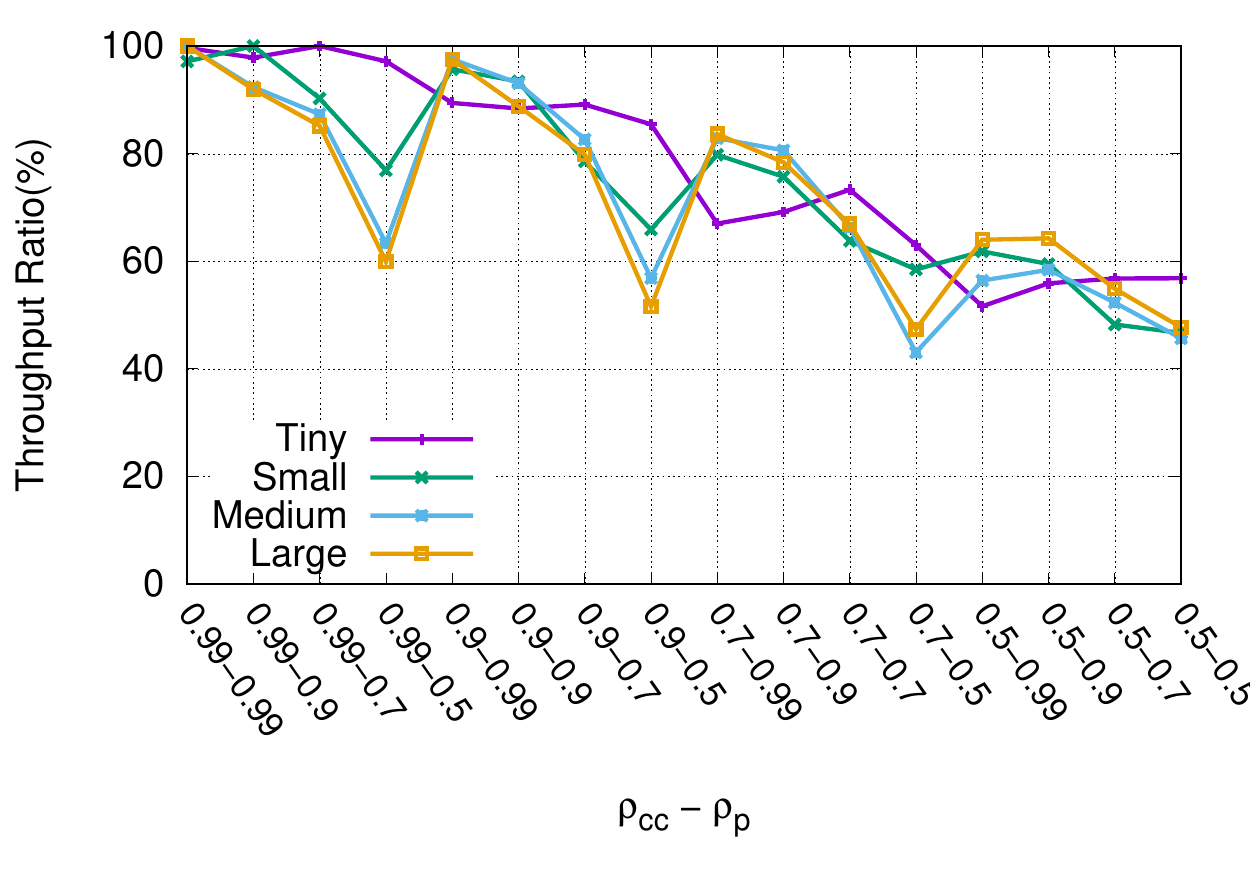}
\includegraphics[keepaspectratio=true,angle=0,width=60mm] {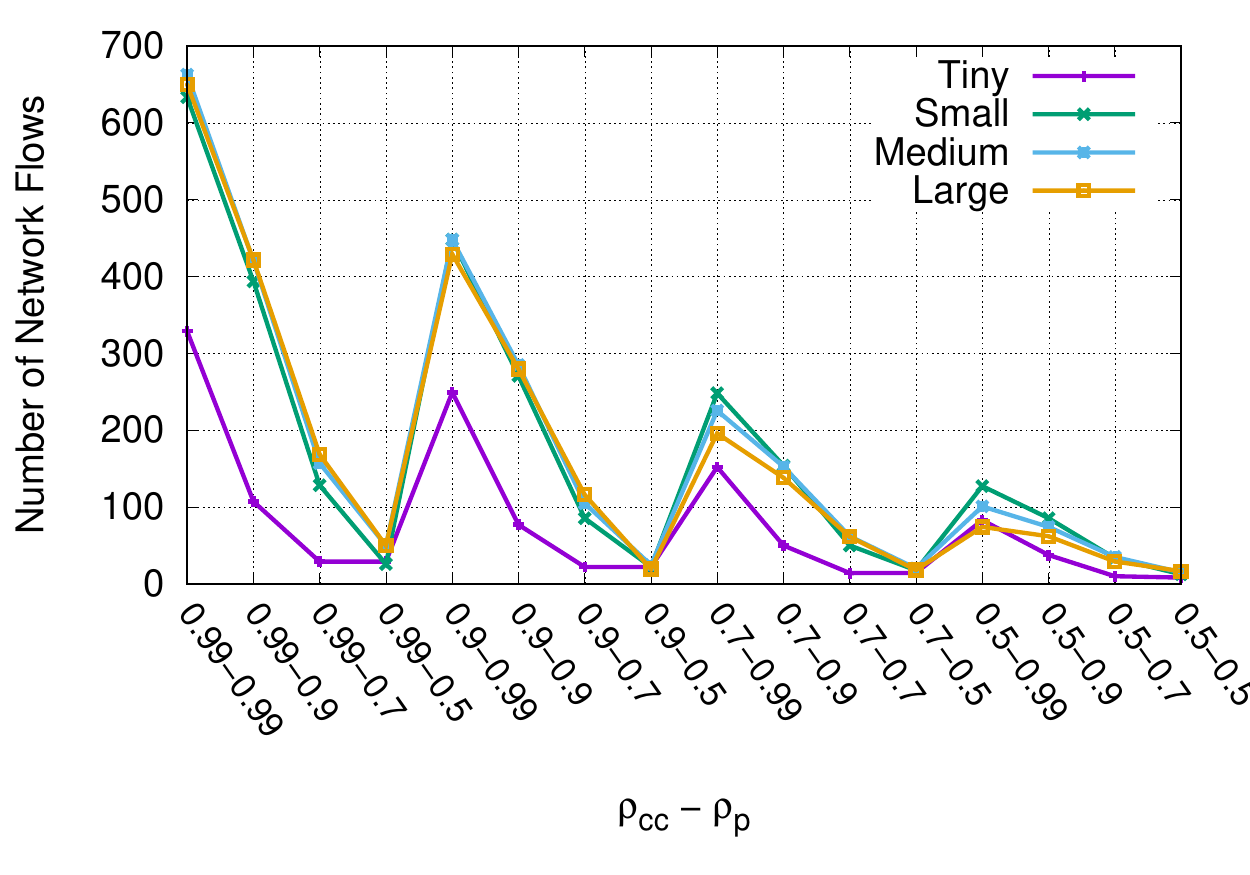}
\caption{Comparison of different relaxation ratios for concurrency-parallelism pair for different file sizes in WAN (XSEDE)} 
\label{fig:relaxation_xsede}
\vspace{-2mm}
\end{figure*}

\begin{figure}
\begin{centering}
\includegraphics[keepaspectratio=true,angle=0,width=72mm] {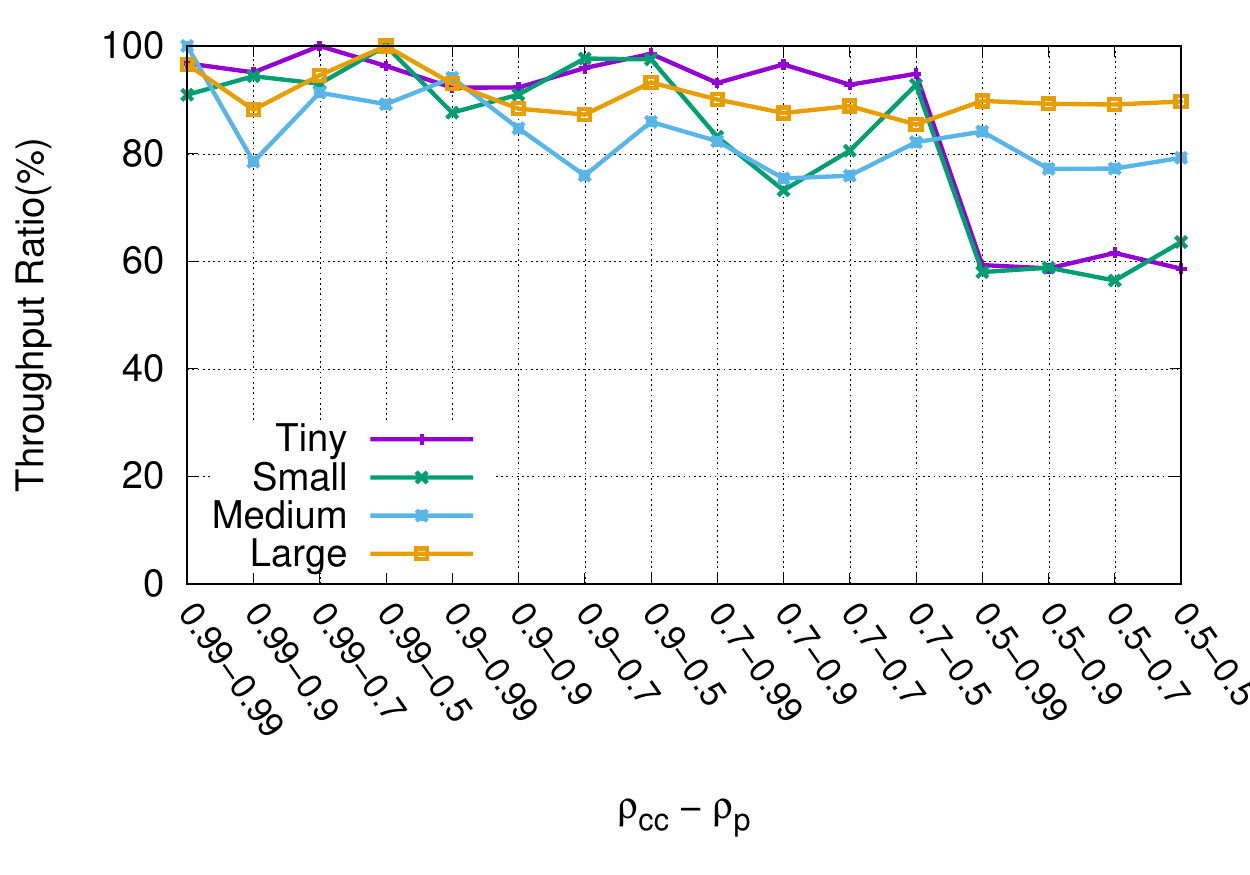}
\caption{ $\rho_p-\rho_{cc}~(0.7-0.7 )$ achieves around 80\% throughput of maximum for all file types in LAN.} 
\label{fig:relaxation_didclab}
\end{centering}
\vspace{-4mm}
\end{figure}

\subsubsection{Variable Relaxation and Combiner} \label{sec:combiner}

Once corresponding parameter values for maximum throughput are found for each $f_i$. $1<i<k$, Optimizer runs relaxation process during which the values of parameters are lowered if change in the estimated throughput stays within a reasonable range. For example, pipelining has either no or little contribution to the throughput on large files, but the optimal point may estimate large pipelining value for a marginal gain. Optimizing cost of energy consumption can be shown as another reason for the importance of relaxation process, as shown in~\cite{power-aware}, that marginal increase in transfer throughput by means of using large values of parameter values may lead to considerable increase in power consumption. 

In the relaxation phase, Optimizer evaluates smaller values for each parameter until the new estimated throughput is larger than certain percentage of the original estimated throughput. Assume $(32, 20,24)$ is calculated as optimal values for concurrency, parallelism and pipelining for an equation $f$ such that $f(32, 20,24) = Tmax$. The relaxation process evaluates smaller values for concurrency starting from 31, while keeping parallelism and pipelining same, until new throughput becomes smaller than certain percentage of initially estimated throughput ($Tmax{'} < \rho * Tmax$). 
We observed that 0.99 relaxation ratio for pipelining works well enough to detect high pipelining value estimation by non-linear programming solver for small throughput gain for large file types. Moreover, using large pipelining values for small file types does not have negative impact on resource utilization, hence on power consumption, since pipelining is only about how many additional transfer command to be stored at source-destination pairs. On the other hand, higher values of concurrency and parallelism mean higher load on end servers and network thus have to be used sparingly. 

We evaluated different values of threshold ($\rho$) for concurrency and parallelism while keeping threshold for pipelining at 0.99 as shown in Figure~\ref{fig:relaxation_xsede}. In the Figures~\ref{fig:relaxation_xsede}(a) and~\ref{fig:relaxation_xsede}(b), y-axis represents ratio of throughput achieved at a relaxation point to maximum throughput achieved among all relaxation values. Although the effect of using small threshold values for parallelism does not seem to be obvious in {\it{Light Traffic}} case, it is easily noticeable in {\it{Heavy Traffic}} case. Moreover, while parallelism has little impact on throughput of {\it{Small}} file size under light background traffic, its impact becomes much more apparent and the throughput decreases considerably as  $\rho_p$ decreases for fixed  $\rho_{cc}$ as shown in Figure~\ref{fig:relaxation_xsede}(b). Figure~\ref{fig:relaxation_xsede}(c) shows the number of network flows opened for different relaxation thresholds. Although disabling relaxation or using large thresholds would gain higher throughput, it would open as much as 650 network flows. In addition to the network overhead, it will create up to 20-30 processes at the end servers which will increase the load at the end servers as well. Thus, we have targeted to find a relaxation threshold that would achieve reasonable throughput without causing too much overhead on network and end systems. Hence, we picked 0.7-0.7 threshold combination for concurrency and parallelism which achieved 60\% or more of maximum throughput while keeping the number of flows less than 80 for all file types. We have also confirmed that 0.7-0.7 threshold combination works well in local area experiments as well, as shown in Figure~\ref{fig:relaxation_didclab}.
 
\begin{equation}
Tmax_i = f_i(cc_i, p_i, pp_i)
\label{eq:maximize}
\vspace{-1mm}
\end{equation}

\vspace{-3mm}
\begin{equation}
\begin{aligned}
cc_{avg} = \sum_{i=1}^{N} \frac{cc_i * w_i}{w_{total}}\hspace{8mm}
p_{avg} = \sum_{i=1}^{N} \frac{p_i * w_i}{w_{total}}\\
pp_{avg} = \sum_{i=1}^{N} \frac{pp_i * w_i}{w_{total}}\hspace{16mm}\\
\end{aligned}
\label{eq:average}
\end{equation}

\subsubsection*{Cost Analysis of~\algoName}\label{sec:cost}
\algoName~runs sample transfers and applies data modeling on the fly so it comes with an overhead. To minimize the overhead, we pipeline the transfer sampling process with optimization process at the best effort. Instead of waiting for each chunk's sample transfer to be completed, we run Optimizer for a chunk as soon as its sample transfer is completed so that Scheduler and Optimizer can operate simultaneously. For example, once Scheduler finishes sample transfer for small chunk, it starts running sample transfer for medium chunk. While sample transfer for medium chunk runs, it passes sample transfer throughput of small chunk to Optimizer so that it runs calculations and returns values for protocol parameters. Since Optimizer can finish the estimation calculations in around 2-3 seconds for each chunk, the bottleneck in the pipelined process becomes the sample transfers. Then, the overall cost boils down to the cost of the sample transfers plus running Optimizer for the last chunk.

\begin{equation}
t_0 = \frac{D}{Thr_0} \label{eq:time-normal}
\end{equation}

\begin{equation}
t_H = \frac{D- D_S}{Thr_H} + \frac{D_S}{Thr_S} + c \label{eq:time-harp}
\end{equation}

\begin{equation}
t_H = \frac{D- (15 * Thr_S)}{Thr_H} +15 + c \label{eq:time-simplified}
\end{equation}

 Equation~\ref{eq:time-normal} shows the duration of the data transfer when \algoName~is not used. $D$ refers to total data size and $Thr_0$ refers to the achieved throughput. When \algoName~is used, the duration of the data transfer is determined by Equation~\ref{eq:time-harp} in which $D_S$ again refers to size of data transferred during sample transfer and $Thr_H$ and $Thr_S$ refers to the throughputs obtained in sample transfer and actual dataset transfers, respectively. $c$ refers to the cost of Optimizer for running the optimization process for the last chunk. As explained in Section~\ref{sec:sample_transfer}, when adaptive sampling approach is used with 5\% threshold and 3 seconds monitoring interval, sampling finishes in less than 15 seconds. Thus, $\frac{D_S}{Thr_S}$ becomes 15, so the sample transfer data size, $D_S$, will be $15 * Thr_S$. Hence $t_H$ reduces to Equation~\ref{eq:time-simplified}.

\begin{table}[t]
\begin{centering}
\begin{tabular}{ |c| c| c| c| |}
\hline
\multirow{2}{*}{\bf  {$T_H$ Speed-up (\%)}} &\multirow{2}{*}{\bf $T_S$ Slowdown (\%)}  & {\bf Min. Chunk}\\
 & &  {\bf Size ($D$) ($\times Thr_0$)}\\
\hline
10 &  50 &  90 \\
\hline
10 &  30 &   60 \\
\hline
10 &  10 &    30 \\
\hline
30 &  50 &   40 \\
\hline                                                                                                                                                                                                                                                                                                                                                                                                                                                                                                                                                                                                                                                                                                                                  
30 &  30 &   30 \\
\hline                                                                                                                                                                                                                                                                                                                                                                                                                                                                                                                                                                                                                                                                                                                                  
30 &  10 &   20 \\
\hline
50 &  50 &    30 \\
\hline                                                                                                                                                                                                                                                                                                                                                                                                                                                                                                                                                                                                                                                                                                                                  
50 &  30 &   24 \\
\hline                                                                                                                                                                                                                                                                                                                                                                                                                                                                                                                                                                                                                                                                                                                                  
50 &  10 &   18 \\
\hline
\end{tabular}
\caption{Minimum chunk size for~\algoName~to pay off.} \label{tab:cost-analysis}
\end{centering}
\vspace{-4mm}
\end{table}

In Table~\ref{tab:cost-analysis}, we calculated minimum data size ($D$) needed for \algoName~to amortize the cost it induces (aka $t_0 = t_H$) under different $Thr_H$ speed-ups and $Thr_S$ slowdowns. {\it $T_H$} Speed-up column represents $\frac{Thr_H-Thr_{0}}{Thr_0}$ which stands for throughput gain when~\algoName~is used. Although the gain will be much higher when \algoName~is compared with Globus Online~\cite{globusonline} and PCP~\cite{esma-tcc}, we compared \algoName~against heuristic (ProMC) we proposed in our earlier work~\cite{europar13} which outperforms Globus Online and PCP by a significant margin. {\it $T_S$} Slowdown column represents the ratio of transfer throughput decrease during sample transfers. Since we transfer each file group (chunk) separately in sample transfers, the obtained throughput is generally smaller than ProMC which transfers multiple chunks simultaneously. For example 30\% slowdown means $Thr_S = (1-0.3) * Thr_0$.  {\it Min Chunk Size} column is represented in $Thr_0$ as $Thr_H$ and $Thr_S$ are evaluated in $Thr_0$ order.

In the worst case scenario, the gain of \algoName~is 10\% and the sample transfers are 50\% slower than $Thr_0$, minimum chunk size that \algoName~pays off the cost is $90*Thr_0$. It will become 67 GB when $Thr_0$ is 6 Gbps. The minimum chunk size to benefit from \algoName~reduces as the throughput gain increases or sample transfer slowdowns reduces. Our observations on the tests we ran in XSEDE, AWS and DIDCLAB networks, we observed that slowdown mostly stays lower than 50\% and the gain ranges from 10\% to 80\%. Hence, the results we present in Section~\ref{sec:evaluation} show that \algoName~mostly outperforms the heuristic algorithms under light traffic and improves the overall throughput significantly under medium and heavy background traffic cases in which case $Thr_H$ speed up reaches up to 50\%.

\section{Experimental Analysis}\label{sec:evaluation}
We compared HARP against heuristic (Globus Online~\cite{globusonline}, Single Chunk, and ProActive Multi-Chunk~\cite{europar13}), probing based~\cite{esma-tcc}, and hysteresis based~\cite{zulkar-ndm14} algorithms. Globus Online (GO) separates the dataset into chunks based on file size and uses predefined values for protocol parameters for each chunk's transfer. Single Chunk (SC) algorithm, similarly, separates the dataset based on file size and transfers them one by one with the protocol parameter values found by heuristic calculations using dataset and network metrics. ProActive Multi-Chunk (ProMC) creates chunks and determines values of protocol parameters similar to SC, but instead of transferring chunks sequentially, it transfers multiple chunks at the same time in order to minimize the effect of small files on overall transfer throughput. Since SC and ProMC require user input for upper bound of concurrency level of a transfer, we have set it to 10 as they seem to be performing the best when maximum concurrency is set to 10~\cite{europar13}. PCP~\cite{esma-tcc} employs a divide-and-transfer approach similar to SC and GO. It determines protocol parameters by running several sample transfers. Finally, ANN+TO models transfer throughput based on historical data and runs sample transfer to learn the current load on the network. Similar to SC and Globus Online, it transfers chunks one by one due to which its overall performance for dataset with mixed data files is dominated by the throughput of small files.

We tested \algoName~both at the networks for which historical data have and have not matching entries in terms of source-destination pairs of data transfers. Datasets used in the experiments are different from the ones used in historical data collection process. While datasets in data collection process are homogeneous (e.g. 10000 of 1 MB files are used for small file types), experiment datasets are generated such that while all file types (small, large etc.) exist. Additionally, file sizes in a file type are determined randomly such that we do not assume any file size distribution within a group as well. Our experiments on XSEDE (from Stampede to Gordon) and DIDCLAB networks are the ones historical data have matching entries and another XSEDE (from Gordon to Stampede) and AWS experiments are the ones historical data does not have entries with same network settings. All the experiments are run at least five times.

\begin{figure}
\begin{center}
\includegraphics[keepaspectratio=true,angle=0,width=72mm] {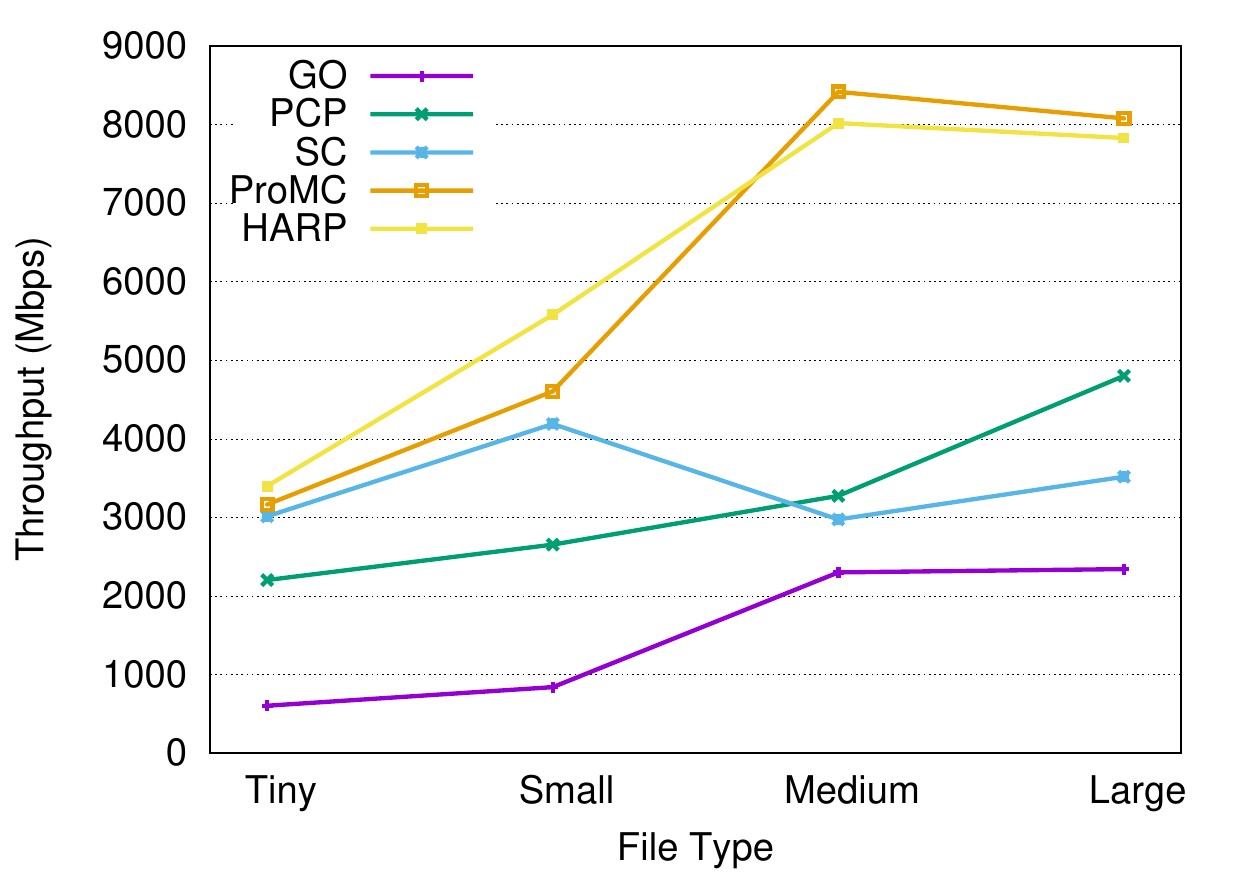}
\caption{Single type file transfers between Stampede (TACC) and Gordon (SDSC) on XSEDE.}
\label{fig:single-type}
\end{center}
\vspace{-6mm}
\end{figure}

We first experimented with transfer of datasets that only have one type of file size; either all large or small files. Figure~\ref{fig:single-type} shows the comparison of GO, ANN+OT, SC, ProMC, and \algoName. Size of datasets are 45 GB and 92 GB for small and large files, respectively. While SC and ProMC achieve similar throughput for small file transfer, they differ in large file transfer. This is because of the way SC calculates concurrency level of a chunk. While ProMC uses all available channels (in this case it is 10), SC may prefer using less number of channels than available. SC determines the number of channels by taking minimum of what it calculates and what is given by user as upper bound. SC calculates more than 10 channels for small files so, it uses concurrency value 10 as we set upper bound to 10. On the other hand, it estimates concurrency value 2 for large files thus, yields lower throughput than ProMC. While ANN+OT performs worse than SC in small file transfers, it outperforms SC by 32\% in large file transfers since it predicts concurrency value larger than what SC calculates. PCP also performs worse than SC for small files which is due to overwhelming effect of sample transfers. Even though it also runs sample transfers for large files, small files are more sensitive to values of concurrency and pipelining than large files. When small values of concurrency and pipelining are used during probing process, it takes long time to finish transfer which affects overall transfer considerably. Moreover, \algoName~outperforms SC and ProMC by around 25\% for small files. However, \algoName~outperforms ProMC only around 5\% for large files. Digging into details, we found out that Optimizer estimates concurrency level in 10-13 range for large files which is close to what ProMC is set to run in this experiment. Although \algoName~yields higher throughput after Optimizer estimates concurrency, due to overhead of sampling and optimization process, its  gain becomes marginal. It is worth to note that while ProMC performs close to \algoName~in this experiment, one has to know what concurrency value to pass to ProMC which requires some degree of knowledge on transfer parameters as well as dataset characteristics.  Finally, \algoName~outperforms ANN+OT for both small and large files which proves that \algoName~does better job in modeling and finding best values of optimal parameters.

Figure~\ref{fig:xsede}
presents the results that we obtained in a networks where historical dataset contains exact matching entries. We have tested algorithms under three different network loads (light, medium, and high background traffic as described in ``Data Collection'' section). In this network, the performance of most algorithms decreased by 50-300\% as network load increases.

\begin{figure*}
\begin{center}
\includegraphics[keepaspectratio=true,angle=0,width=60mm] {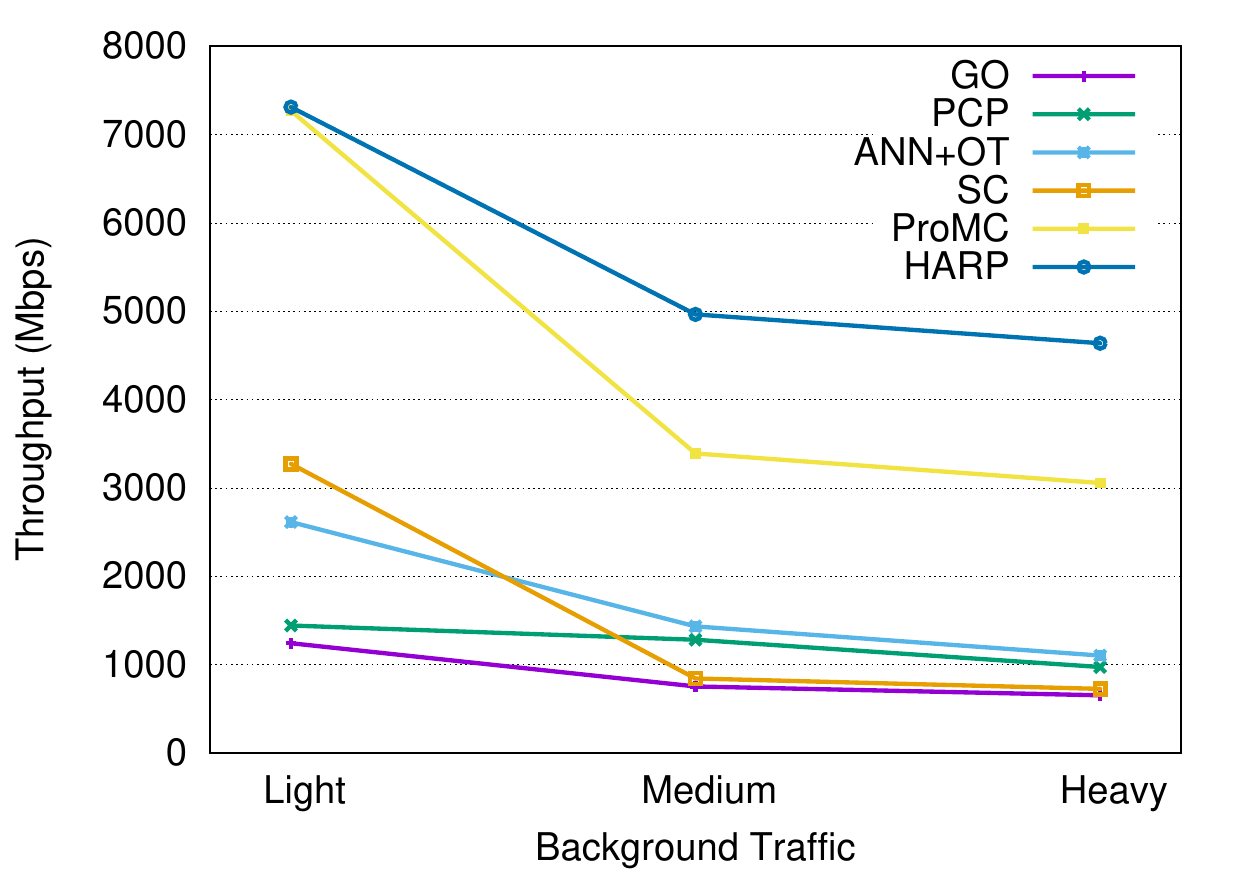}
\includegraphics[keepaspectratio=true,angle=0,width=60mm] {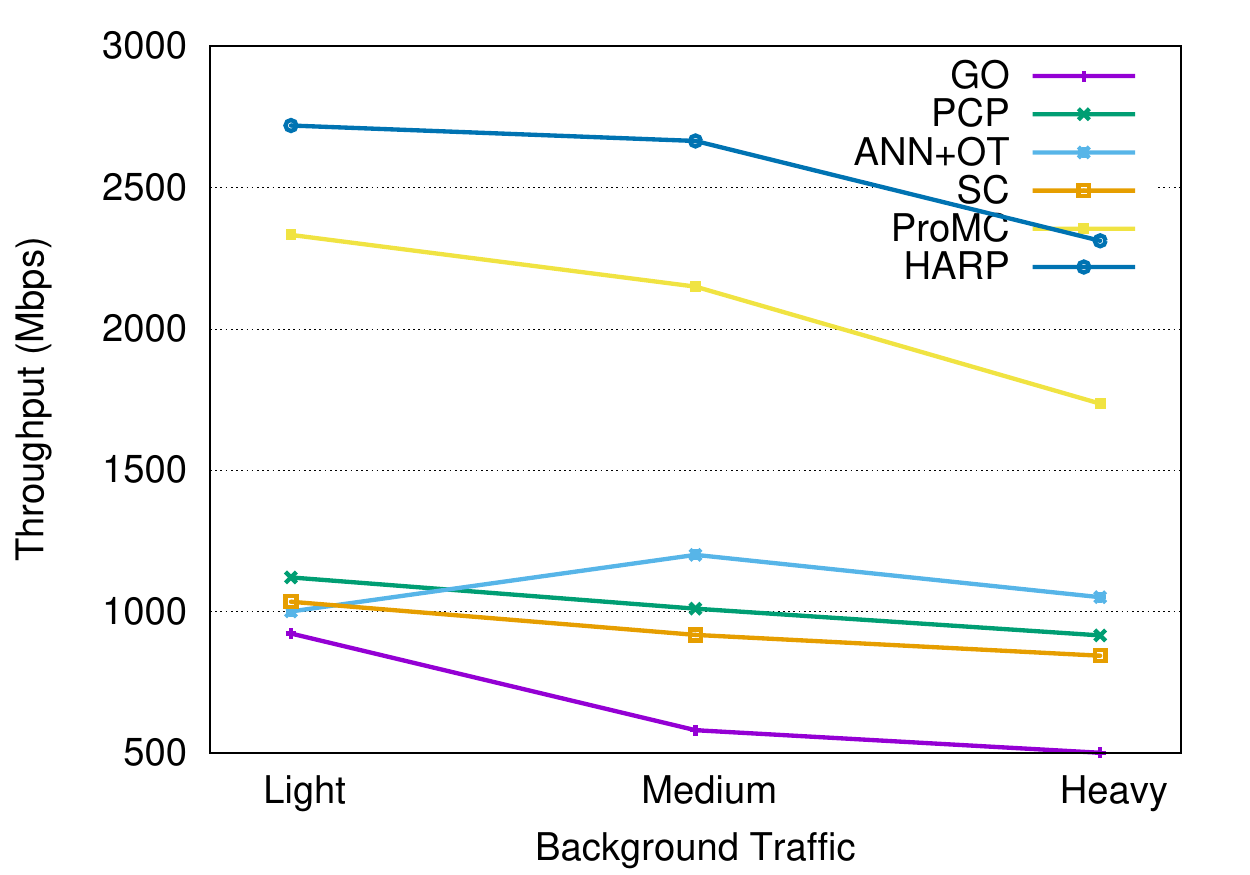}
\includegraphics[keepaspectratio=true,angle=0,width=60mm] {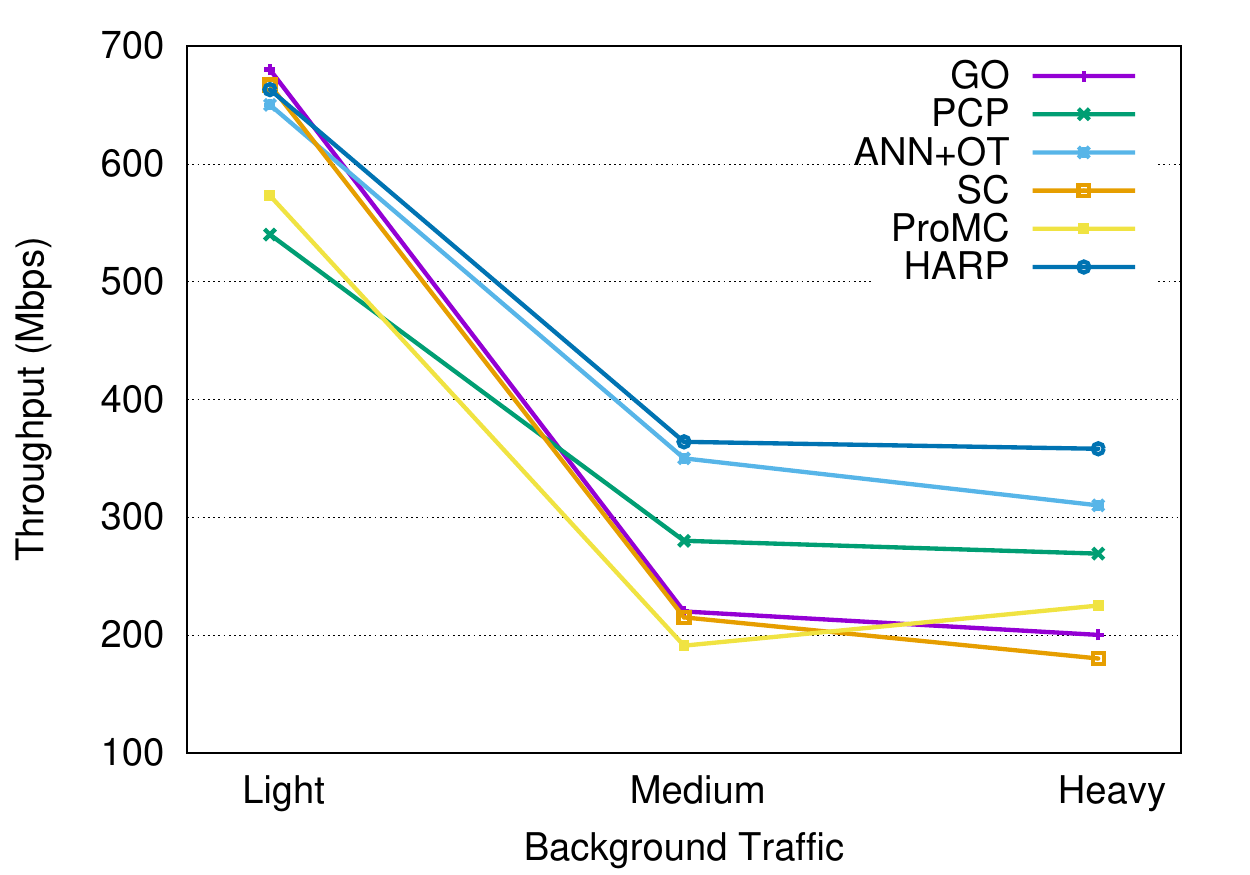}
\caption{HARP adapts protocol parameters to varying background traffic and obtains higher transfer throughput.}
\label{fig:instantaneous}
\end{center}
\vspace{-6mm}
\end{figure*}

Algorithms that transfer one chunk at a time (GO, ANN+OT, PCP, and SC) exhibit poor performance in XSEDE because their overall performance is pulled down by the throughput of small file transfers. GO, ANN+TO, PCP, and SC achieve less than 3 Gbps under light background traffic. On the other side, multi-chunk algorithms (ProMC, and \algoName) are able to deliver around 7 Gbps. As discussed in Section~\ref{sec:cost}, \algoName-requires data size to be greater than certain amount to outperform the heuristics. The size of dataset used in Figure~\ref{fig:xsede} was 130 GB which is only enough to cover the overhead imposed by \algoName. When dataset size is increased to 260 GB, throughout of \algoName~reached to 8 Gbps as \algoName's~overhead is alleviated by the increase in overall transfer duration. 
Unlike light background traffic case, \algoName~gains 36\% and 48\% more throughput than ProMC even for smaller dataset size (130GB) under medium and heavy background traffic cases by taking advantage of historical transfer information. The reason why ProMC performs worse as the network load varies is its traffic-agnostic parameter estimation approach. One may claim that ProMC can learn about network load by performing sample transfers. However, without having historical information, it cannot interpret probing results. A simple way to interpret probing is done by PCP algorithm which runs several probings and increments parameter values until probing throughput decreases. However, results show that it fails to capture high overall transfer throughput as the number of probings becomes high to accurately identify ``right value'' for parameters.
The throughput of SC algorithm is dropped drastically from 3.2 Gbps to 750 Mbps as network load increases. Similarly, throughput of GO suffers significantly as network traffic increases. ANN+OT and PCP are able to adapt protocol parameters accordingly and outperform SC and GO under high network loads. However, overhead of sampling and ``one chunk at a time'' policy limits their overall performance to less than 1 Gbps in heavier network loads.

Since heuristic algorithms are unaware of disk subsystems when calculating  protocol parameters, they pick parameter values solely based on network and end system configurations. However, this simple heuristic approach might be misleading when transfer throughput is limited by disk I/O performance and disk I/O throughput decreases as the number of active threads increases. Hence, \algoName~can outperform heuristic algorithms even when relatively small dataset with the help of historical data. When there is no background traffic, \algoName~achieves 13\% higher throughput than ProMC. As the network load increases, \algoName~achieves 47\% and 37\% more than ProMC for medium and high network load experiments. ANN+OT outperforms heuristics under all network loads but falls short to compete with \algoName~due to single chunk approach.

In order to test the effectiveness of \algoName~for networks that have no matching entries in historical data, we run experiments on XSEDE and Amazon EC-2. Although same pair of servers are used in XSEDE experiments, we have transferred dataset in a reverse path of historical data entries. Namely, historical data has the logs for transfer that are sourced from Stampede and destined to Gordon. In this experiment, Gordon is used as a source and Stampede became the destination. Although it may seem to be identical of Stampede-Gordon transfers, the results shows us that the maximum achievable throughput is different than Stampede-Gordon transfers basically due to (i) end system storage speeds are different for read and write operations, (ii) at any given time free network bandwidth is less than what is observed in reverse direction. Hence, Gordon-Stampede is a good example to see how \algoName~performs on the networks that have a similar but not exact entries in historical data.

As opposed to Stampede-Gordon transfers, Gordon-Stampede transfers are more disk I/O bound which can be deduced by looking at throughput change as the background traffic increases. While throughput decrease ranges in 50-300\% in Stampede-Gordon transfers as network load increases, it stayed around 10-100\% in Gordon-Stampede transfers. \algoName, ANN+OT, and PCP are affected the least by increased network traffic compared to heuristics since they probe network status at the beginning of transfer and picks parameter values accordingly. For example, \algoName~gained 13\% more throughput than ProMC under light background traffic. The improvement ratio increased to 24\% as throughput of ProMC dropped by 24\% under heavy network throughput while throughput of \algoName~only dropped by 11\%. Moreover, the difference between maximum observed throughput and throughput obtained by \algoName~increased when compared to earlier experiments. This is an expected behavior since Optimizer selects logs of Stampede-Gordon transfers during modeling phase and optimal values for protocol metrics for Gordon-Stampede transfers are not the best ones for Stampede-Gordon transfers as explained in Section~\ref{sec:motivation}.

\begin{figure*}
\centering
\begin{minipage}{.46\textwidth}
        \centering
        \includegraphics[keepaspectratio=true,angle=0,width=65mm] {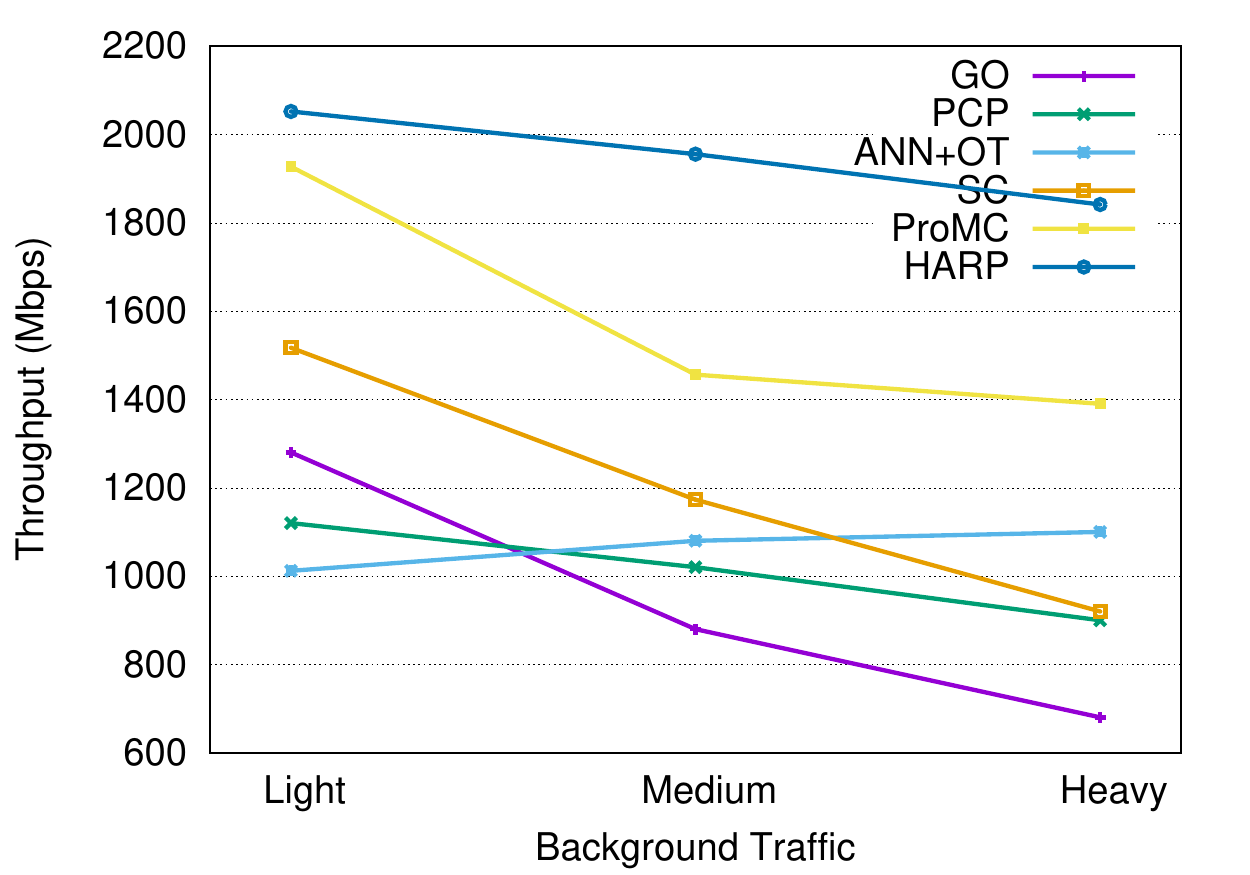}
        \caption{\algoName~outperforms ProMC in all traffic types in Cloud experiments. The difference goes up to 30\% under heavy background traffic.}
        \label{fig:aws}
    \end{minipage}%
    \hspace {.07\textwidth}
    \begin{minipage}{0.46\textwidth}
        \centering
        \includegraphics[keepaspectratio=true,angle=0,width=65mm] {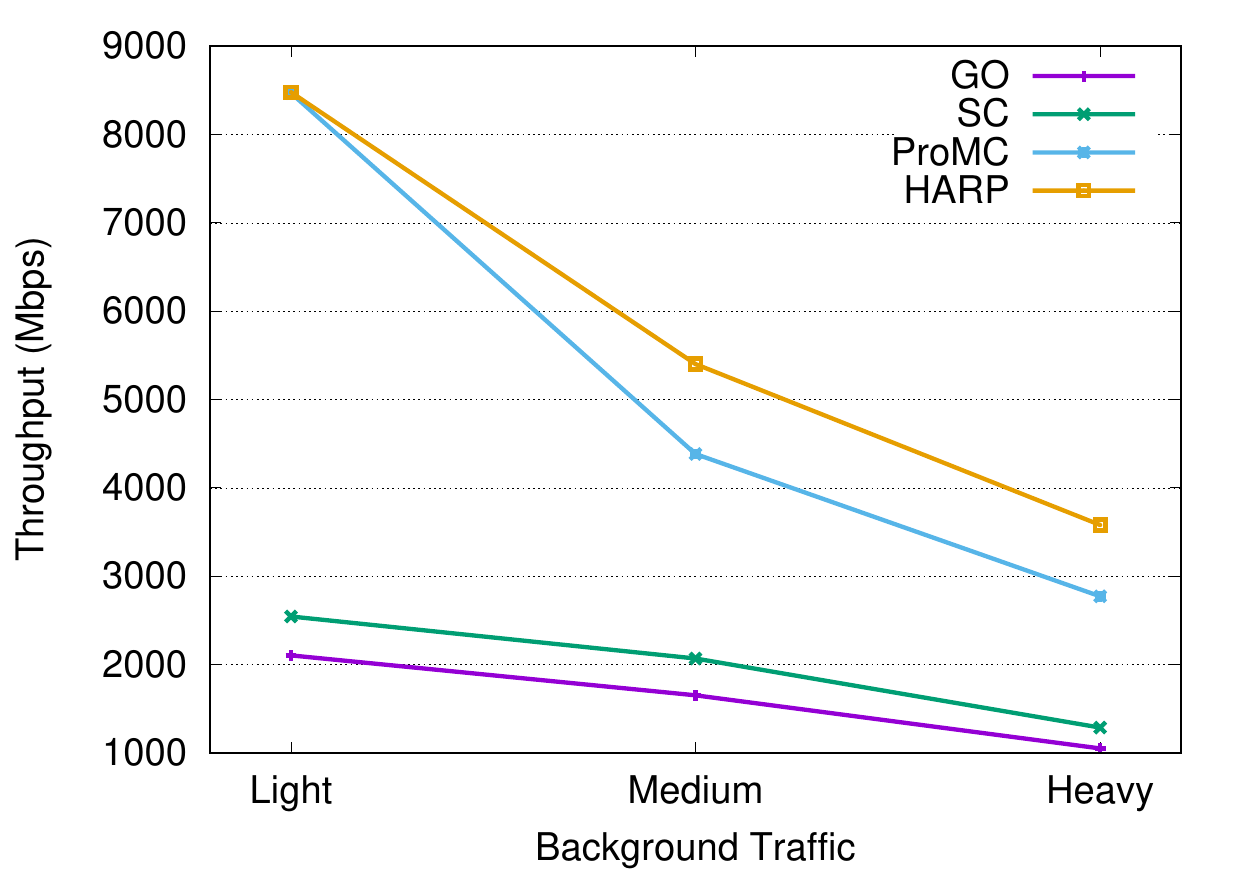}
        \caption{\algoName~achieves 23\% to 30\% speed-up over ProMC in Dark Energy Survey data transfer under medium and heavy background traffic.}
        \label{fig:dark-energy}
    \end{minipage}
\end{figure*}

Finally, we tested \algoName~in Amazon EC-2 for which we used two c3.8xlarge instances with network specification given in Table~\ref{tab:system-spec}. We have used Provisioned IOPS EBS storage volume as it offers the highest disk I/O throughput (around 320 MB/s). However, we were able to achieve around 265 MB/s disk throughput because of file operation overheads as a result of having too many small files. Similar to XSEDE experiments, algorithms that transfer multiple chunks simultaneously (ProMC and \algoName) performed better than single chunk algorithms at all background traffic loads as shown in Figure~\ref{fig:aws}. Similar to the Gordon-Stampede experiments, transfer throughputs are not as much affected as Stampede-Gordon experiments by increased network load since transfer throughputs are again limited by disk I/O throughput.

Due to similarities in network settings, cosine-similarity favors WAN transfer logs over LAN transfer logs when filtering similar entries in historical data. Since EC-2 network resembles to XSEDE network in context of yielding higher disk I/O throughput as the number of concurrent transfer increases and having a smaller buffer size than BDP, XSEDE entries-based derived model works well in EC-2 experiments. Hence, \algoName~outperforms ProMC by 5\% in light background traffic case. The difference reaches to 34\% and 32\% under medium and high network loads as ProMC fails to adapt protocol parameters to varying system load.

\subsection*{Optimizing Dark Energy Survey Data Transfer}
Dark Energy Survey\cite{des} captures pictures of space to probe the dynamics of the expansion of the Universe and the growth of large-scale structure. It collects 200-300 GB of data each day that is transferred from observatory in Chile to collaborating research institutions in America and Europe.
We took one day worth of data which consists of 428 files, sizes range from 270 MB to 730 MB, and transferred from Stampede (TACC) and Gordon (SDSC) under different background traffic.

Figure~\ref{fig:dark-energy} shows comparison of Globus Online, heuristics (SC and ProMC), and~\algoName~in transferring Dark Energy Survey data. SC and Globus Online performs 2-3 times less throughput than ProMC and~\algoName~due to underestimation of protocol parameters. ProMC and \algoName~performs similar under light background traffic as they both estimate to use similar parameters and obtain close-to-maximum transfer throughput. However, as background traffic increases ProMC is affected significantly and falls behind~\algoName by 23\% and 30\% due to failing to adapt protocol parameter values to the changing background traffic.

\subsection{Accuracy of the Model}

\begin{table*}[t]
\begin{centering}
\begin{tabular}{ |p{1.7cm}|p{4mm}|p{7.3mm}|p{4mm}|p{7.3mm}|p{4mm}|p{7.3mm}|p{4mm}|p{7.3mm}|p{4mm}|p{7.3mm}|p{4mm}|p{7.3mm}|p{4mm}|p{7.3mm}|p{4mm}|p{7.3mm}|}
\hline
 &\multicolumn{8} {c|} {Stampede-Gordon (WAN)} & \multicolumn{8}{c|} {WS1-WS2 (LAN)}\\
 \hline
{\bf File Type }&\multicolumn{2} {c|} {Tiny} & \multicolumn{2}{c} {Small} & \multicolumn{2}{|c|} {Medium} & \multicolumn{2}{c|} {Large}&\multicolumn{2} {c|} {Tiny} & \multicolumn{2}{c} {Small} & \multicolumn{2}{|c|} {Medium} & \multicolumn{2}{c|} {Large}\\
\hline
{\bf Traffic}&Light& Medium&Light& Medium&Light& Medium&Light& Medium&Light& Medium&Light& Medium&Light& Medium&Light& Medium\\
\hline
{\bf Concurrency}&  24 & 25  & 22 & 22 & 10 & 10  & 12 & 10 & 6& 4& 2& 1& 1& 1& 2& 1\\
\hline
{\bf Parallelism} & 0 & 0 & 11 & 10 & 18 & 19 & 15 & 19 & 3& 6& 7& 11&9& 13& 10 & 9\\
\hline
{\bf Pipelining}& 5.5 & 4 & 0 & 0 & 0 & 0 & 0 & 1 & 2& 1& 1& 2& 1& 2 & 2 & 1\\
\hline
{\bf Validation}&  \multirow{2}{*} {95} &  \multirow{2}{*} {96} &  \multirow{2}{*} {94} & \multirow{2}{*} {93} & \multirow{2}{*} {90} &  \multirow{2}{*} {85} &  \multirow{2}{*} {93} & \multirow{2}{*} {91}&
  \multirow{2}{*} {90} &  \multirow{2}{*} {86} &  \multirow{2}{*} {93} & \multirow{2}{*} {91} & \multirow{2}{*} {90} &  \multirow{2}{*} {88} &  \multirow{2}{*} {88} & \multirow{2}{*} {88}\\
{\bf Accuracy (\%)} & & & & & & & & & & & & & & & &\\
\hline
{\bf Estimation} &  \multirow{2}{*} {41} &  \multirow{2}{*} {62} &  \multirow{2}{*} {78} &  \multirow{2}{*} {77} &  \multirow{2}{*} {81} &  \multirow{2}{*} {73} &  \multirow{2}{*} {85} &  \multirow{2}{*} {86}&  
\multirow{2}{*} {84} &  \multirow{2}{*} {79} &  \multirow{2}{*} {91} &  \multirow{2}{*} {76} &  \multirow{2}{*} {90} &  \multirow{2}{*} {91} &  \multirow{2}{*} {87} &  \multirow{2}{*} {86}\\
{\bf Accuracy (\%)} & & & & & & & & & & & & & & & &\\
\hline
\end{tabular}
\caption{Sample transfer metrics and accuracy values from \algoName's~Optimizer} \label{tab:model-accuracy}
\end{centering}
\vspace{-4mm}
\end{table*}

Table~\ref{tab:model-accuracy} shows values of the parameters for transfers in Wide Area and Local Area networks under different background traffic cases.
Since transfer parameters have different impacts on different file sizes, we have gathered transfer parameters for each file type separately.

Optimizer is able to differentiate WAN and LAN transfers by picking high concurrency values for WAN transfers and low concurrency values for LAN transfers. For different file size in WAN and LAN transfers, it calculates high concurrency values for small file types and high parallelism values for large file types. Although the optimal values for concurrency and parallelism might be determined as 32 in Nonlinear Equation Solver, Relaxation process decreases them a bit to avoid overloading network and end systems. In addition, it mostly picks higher parallelism values as network becomes more congested in order to receive higher share in network resources.

After {\it Filtering} process of Optimizer selects similar entries from historical data and {\it Grouping} categorizes them, we allot 30\% of each groups as test data and use the rest to train the model. To measure correctness of the derived model, we first calculate optimal parameter values using training data. Let's say it returns ($cc_{training}, p_{training}, pp_{training}$) for corresponding throughput $Thr_{training}$. Then, we use test data and apply polynomial regression to derive model, $f_{test}$, and find optimal parameter values, ($cc_{test}, p_{test}, pp_{test}$), and estimated throughput, $Thr_{test}$. Then, instead of directly comparing throughputs $Thr_{training}$ and $Thr_{test}$, we calculated throughput $Thr_{projected} = f_test(cc_{training}, p_{training}, pp_{training})$ in order to project how close parameters of training data are to the optimal parameters of test data. Direct throughput comparison may not be accurate because test data and training data might have been exposed to different background traffic, thus maximum throughput of two sets of data might be different even if optimal parameters are same. So, to measure correctness of regression analysis, we calculated validation accuracy as $\frac{\abs{Thr_{test} - Thr_{projected}}}{Thr_{projected}}$. We also listed estimation accuracy which measures closeness of estimated throughput to the actual throughput. Estimation accuracy might not be a good metric to judge the model since actual throughput of a network may change over time even though optimal parameters stays same.

Validation accuracy of \algoName~is always above 85\% which indicates success of regression analysis in modelling transfer throughput. While accuracy of throughput estimation in LAN is more stable and comparatively high,  it is worse in WAN experiments since it is an uncontrolled environment so resource capacities might have changed between data collection and experimenting periods. Looking into deeper why estimation accuracy is 41\% for Tiny file type in WAN experiment, we have discovered that while maximum throughput of Tiny files in historical data never reaches beyond 4 Gbps, we have observed 5.5 Gbps in test experiments. Thus, the accuracy of throughput estimation highly depends on consistency of historical data with current network status. This can easily be handled by logging every real time transfer so that historical data can hold up-to-date information.

\subsection{Online Tuning} \label{sec:onlinetuning}

Since transfers can take tens of minutes, hours or even days, it is inevitable that background traffic changes while transfer is running. Hence, one-time sampling at the beginning of the transfer will not work well for long running transfers. Therefore, we extended \algoName~with Online Tuning which periodically monitors transfer throughput and calculates new parameter values based on observed background traffic. 

Online Tuning also helps to eliminate the need for special sampling phase since we can start with some initial values until new tuning parameters are calculated by Optimizer. It saves time from connection startup/tear down cost that sample transfers induces. Also, Scheduler does not also have to wait for Optimizer to finish its operation as they can work simultaneously. Scheduler measures transfer throughput in certain monitor intervals and passes observed throughput and currently used parameter values to Optimizer. Optimizer then can execute modeling and parameter estimation operations described in Section~\ref{sec:optimizer} again to estimate new values for parameters. Meanwhile, Scheduler does not have to wait for Optimizer as the result of Optimizer can be applied in the next interval. That is, while Scheduler is in interval $MI_{i+1}$, Optimizer can calculate new parameters based previous interval $MI_{1}$ and newly proposed values can be applied at the end of current interval. 

Since it is possible that some throughput variations may happen even when no significant background traffic changes happen, we consider last $k$ monitor interval when making decision on whether or not to change parameter values. After at least $k$ periods have passed and Optimizer consistently suggests to use different parameter values then, Scheduler updates parameter values. Among concurrency, parallelism, and pipelining; pipelining is the easiest parameter to change the value as it does not require connection establishment/tear down. On the other hand, concurrency means creating a connection between source and destination which may take couple of seconds due delay and slow authentication process. While parallelism should not require a new connection establishment, current implementation of GridFTP only allows parallelism level to be set when connection is first established. Hence, we have to close an existing connection and establish new connection with an updated parallelism value. Since connection establishment/tear down is a costly operation, we update concurrency and parallelism levels only if there is at least two difference between old and new values such that expected gain pays off the induced cost. For example, if we are currently running transfer with concurrency level 4 and Optimizer suggests to use 5 in the last $k$ periods then Scheduler does not apply it. While the right value for $k$ may affect accuracy and stability of Online Tuning, we observed that $k=4$ works well in the experiments as it is large enough to avoid from instant throughput variations and small enough to catch prolonged background traffic change.

\begin{figure*}
\begin{center}
\includegraphics[keepaspectratio=true,angle=0,width=88mm] {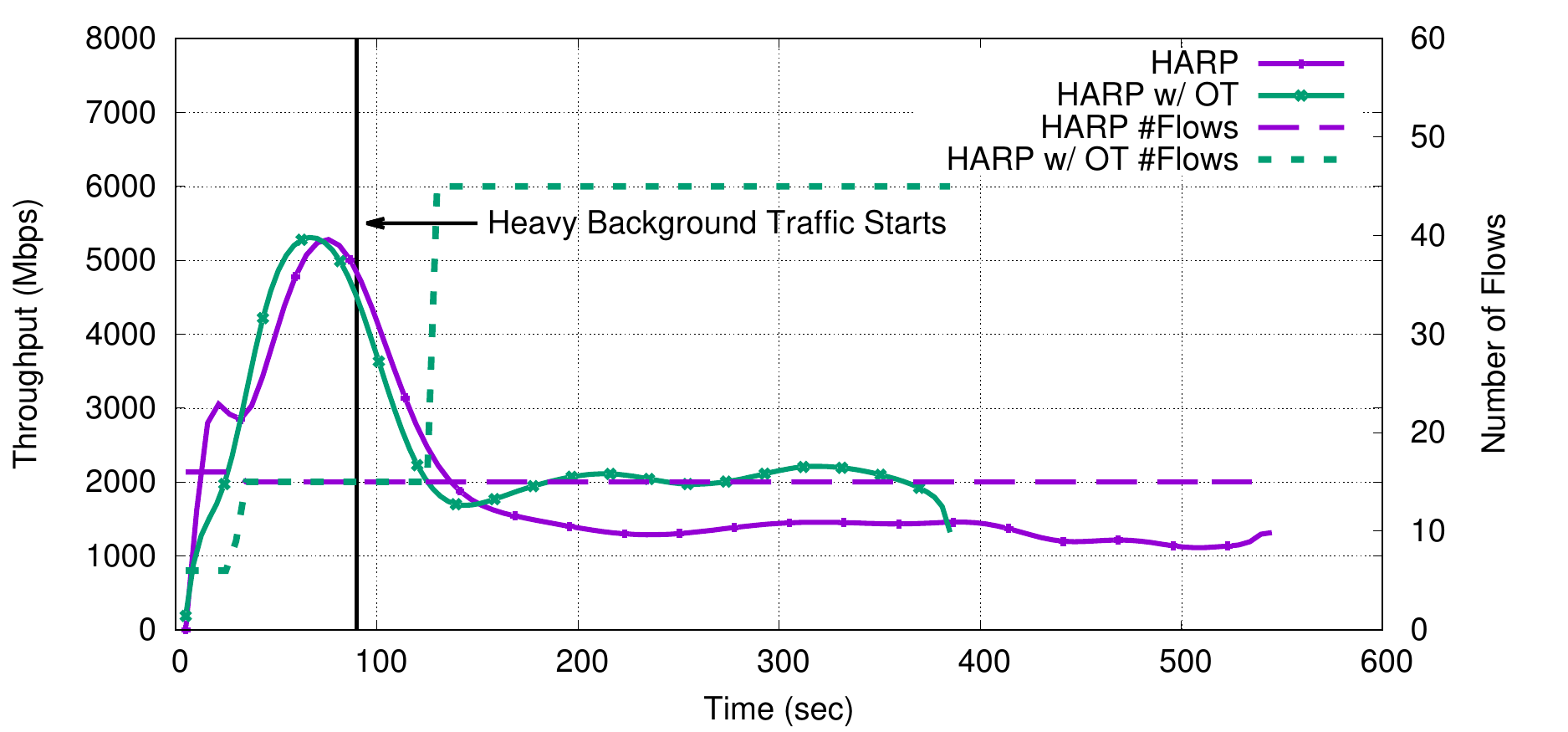}
\includegraphics[keepaspectratio=true,angle=0,width=88mm] {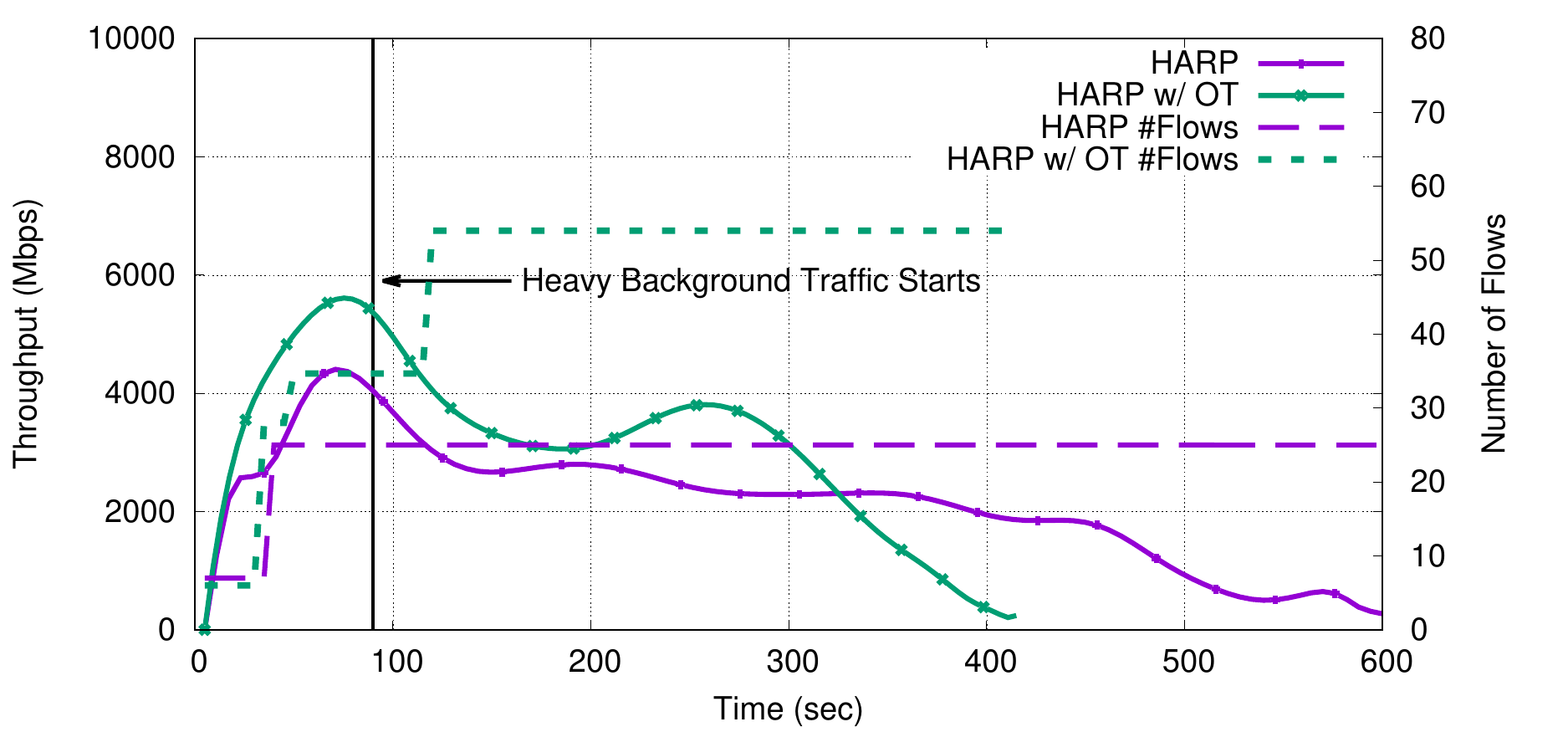}
\caption{Online Tuning detects increasing background traffic and increases the number of flows for higher throughput}
\label{fig:instantaneous}
\end{center}
\vspace{-4mm}
\end{figure*}

\begin{figure*}
\begin{center}
\includegraphics[keepaspectratio=true,angle=0,width=88mm] {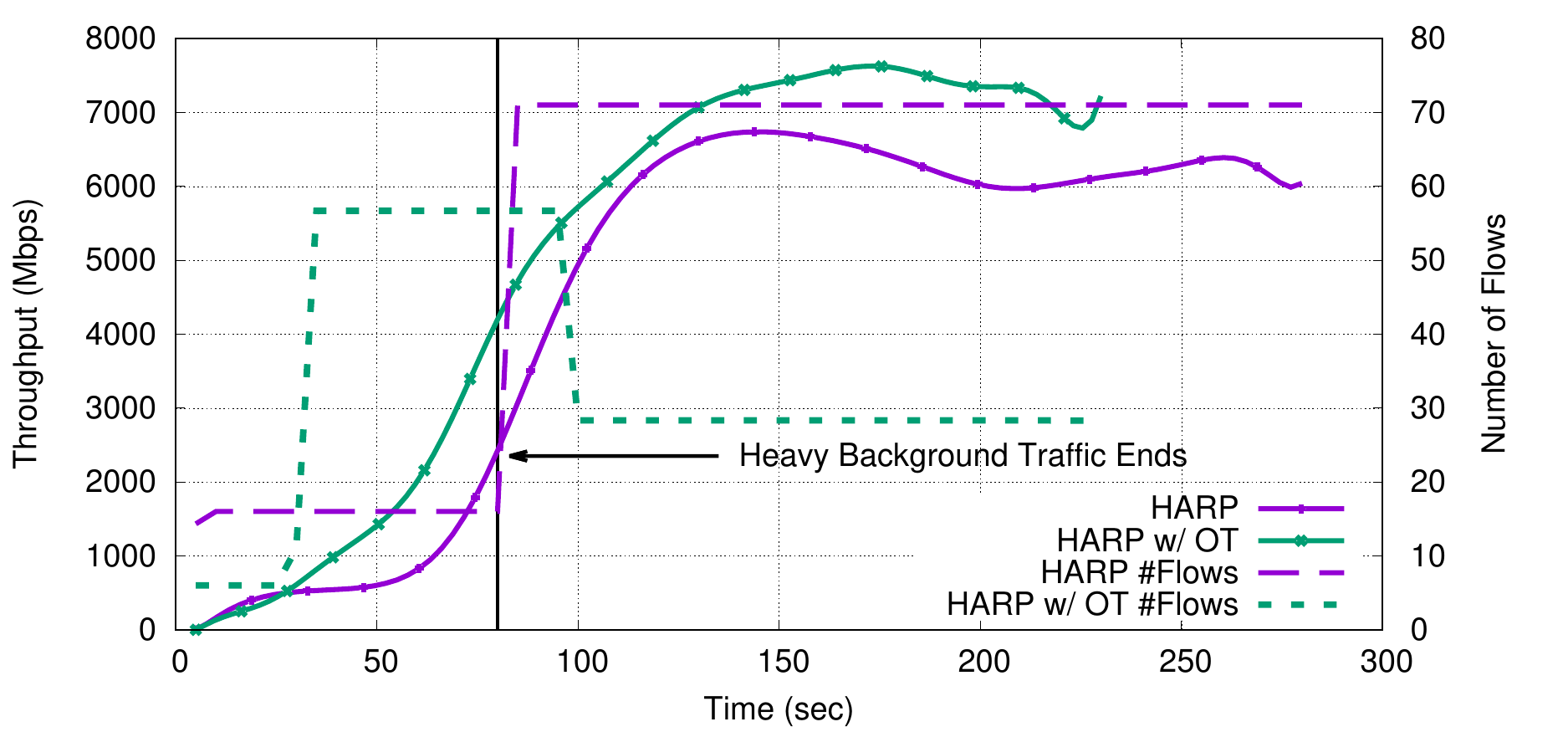}
\includegraphics[keepaspectratio=true,angle=0,width=88mm] {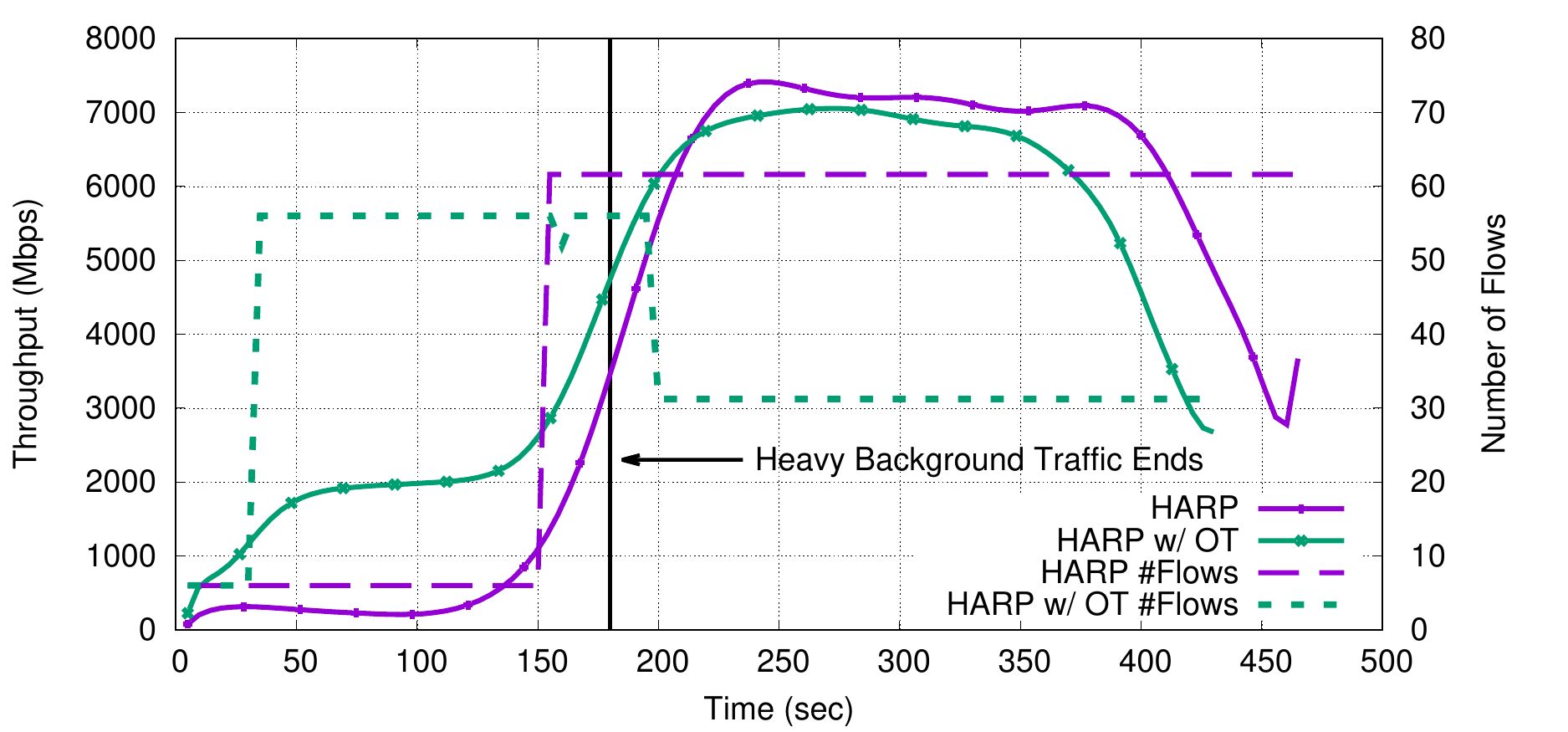}
\caption{Online Tuning detects decreasing background traffic and decreases the number of flows to minimize network overhead}
\label{fig:instantaneous-reducing}
\end{center}
\vspace{-4mm}
\end{figure*}

Figure~\ref{fig:instantaneous} and~\ref{fig:instantaneous-reducing} show comparison of instantaneous throughput of \algoName~and \algoName~with Online Tuning (\algoName~w/ OT) for transfers in XSEDE network. We have tested with Small and Large file types only for the sake of simplicity. We evaluated under dynamic network conditions and transitioned background traffic from: (i) light to heavy (Figure~\ref{fig:instantaneous}), (ii) heavy to light (Figure~\ref{fig:instantaneous-reducing}) in the middle of transfers. Transitions are marked with a solid vertical line.  ``Heavy Background Traffic Starts'' and ``Heavy Background  Traffic Ends'' refers to start and finish of heavy background traffic which is controlled manually. We also tracked the number of flows in the experiments to have a clue about currently used parameter values. Number of flows is calculated by multiplying concurrency and parallelism values and refers to the number of network flows in a given transfer. 

Although we have smoothened instant throughput, it can be seen in the Figure~\ref{fig:instantaneous}(a) that \algoName's~throughput decreases a bit around 30th second. This is because of the the way \algoName~is designed to operate. It first runs sample transfer and passes the results of sample transfer to Optimizer and waits until Optimizer returns. This idle time can be eliminated with the help of Online Tuning since we do not have to have separate sampling phase from actual transfer anymore. Rather, Optimizer and Scheduler can work simultaneously and Scheduler can apply changes in the next interval.

Since Online Tuning requires at least four consecutive intervals of consistent input from Optimizer to apply parameter change, there is some delay between transfer throughput starts decreasing and Online Tuning reacts as in Figure~\ref{fig:instantaneous}(a) and~\ref{fig:instantaneous}(b). While Scheduler does not to require same estimations by Optimizer in four consecutive periods, it expects to receive consistently large or small values for a parameters in order to make sure that higher or lower values offered by Optimizer is not due to transient traffic.

\algoName~without OT estimates parameter values at the beginning of the transfer and keep same value throughout the transfer as Number of Flows stays same. On the other hand when OT is enabled, it can react to varying background traffic and uses higher concurrency/parallelism value to obtain higher transfer throughput as shown in Figure~\ref{fig:instantaneous}. By adapting parameter values to varying background traffic \algoName~with OT achieves 30-40\% higher overall throughput and finishes 160 to 185 seconds earlier.

When \algoName~starts to run when background traffic is heavy, it estimates higher parameter values to obtain high transfer throughput as shown in Figure~\ref{fig:instantaneous-reducing}. However, when background traffic decreases, higher parameter values contributes to throughput slightly if not degrades. For example, higher number of flows leads to achieve around 2-5\% more throughput for large files (Figure~\ref{fig:instantaneous-reducing}(b)) while it causes ~15\% less throughput for small files (Figure~\ref{fig:instantaneous-reducing}(a)). Decreasing the number of flows as background traffic changes from heavy to light leads to higher instantaneous transfer throughput for small files. This is because the level of parallelism used in heavy and light background traffic. When background traffic is heavy, larger parallelism value helps to achieve higher transfer throughput by increasing its share in the network bandwidth. On the contrary, it causes a decrease in throughput when background traffic is low since separating small files into smaller pieces does not help when achieved throughput is already high. Thus, \algoName~with OT outperforms \algoName~even by using less number of flows. As a result, Online parameter Tuning does not only help to keep the number of flows small in return for small performance sacrifice for large files but also obtain higher throughput by using smaller parallelism level for small files.

\section{Related Work}
Liu et al.~\cite{R_Liu10} developed a tool which optimizes multi-file transfers by opening multiple GridFTP threads.
The tool increases the number of concurrent flows up to the point where the transfer performance degrades.
Their work only focuses on concurrent file transfers, and other transfer parameters are not considered.
%

Globus Online~\cite{globusonline}
offers fire-and-forget file transfers through thin clients over the Internet.
The developers mention that they set the pipelining, parallelism, and concurrency parameters to specific values for three different file sizes (i.e. less than 50MB, larger than 250MB, and in between).
However, the protocol tuning Globus Online performs is non-adaptive; it does not consider real-time background traffic conditions. 
Other Managed File Transfer (MFT) systems were proposed which used a subset of these parameters in an effort to improve the end-to-end data transfer throughput~\cite{WORLDS_2004, ScienceCloud_2013, Royal_2011, IGI_2012}.


Other approaches aim to improve the transfer throughput by opening flows over multiple paths between end-systems~\cite{Khanna:2008:UOE:1413370.1413418},
however there are cases where individual data flows fail to achieve optimal throughput because of the end-system bottlenecks.
Several others propose solutions that improve utilization of a single path by means of parallel streams~\cite{Hacker:2005:ADB:1203492.1318153,R_Dinda05, Esma-DADC09},
pipelining~\cite{R_Bres07},
and concurrent transfers~\cite{kosar04, Kosar09}.
Although using parallelism, pipelining, and concurrency may improve throughput in certain cases, an optimization algorithm
should also consider system configuration, since the end-systems may present factors (e.g., low disk I/O speeds or over-tasked CPUs) which can introduce bottlenecks.

Yildirim et al.~\cite{R_Yildirim11}, Yin et al.~\cite{R_Yin11}, and Kim et al.~\cite{KimYK15} proposed highly-accurate predictive models solely based on real-time probing which would require as few as three sampling points to provide very accurate predictions for the parallel stream number giving the highest transfer throughput. These models have proved to provide higher accuracy compared to existing similar models in the literature~\cite{R_Hacker02, R_Dinda05}. 

Later, Yildirim et al. presented the PCP algorithm to dynamically tune parameter values of data transfer~\cite{esma-tcc}. PCP categorizes files in dataset into three groups based on file size (small, medium, and large) and then run sample transfer for each file group to determine parameter values that would return higher transfer throughput. Series of sample transfers are run to determine so-called optimal value of a parameter. Although PCP does not require historical data to operate and it can adapt itself to varying network conditions, too many sample transfers are required to determine ``optimal'' value. Even though original dataset is used during sample transfers, overall transfer throughput are affected by sample transfer throughputs a lot as shown in Evaluation section. 


%
In our earlier work we have proposed heuristic algorithms~\cite{europar13} to determine the best parameter combination by using network and dataset characteristics (i.e bandwidth, round-trip-time, and average file size etc.). Nine et al. developed ANN+OT~\cite{zulkar-ndm14} which uses historical data to derive model that relates transfer metrics to transfer throughput. It then runs real time probing in order to capture current network status.

\section{Conclusions}
In this paper, we presented predictive end-to-end data transfer optimization algorithms based on historical data analysis and real-time background traffic probing, called HARP. Most of the existing work in this area is solely based on real time network probing, which either cause too much sampling overhead or fail to accurately predict the correct transfer parameters. Combining historical data analysis with real time sampling enables \algoName~to tune the application level data transfer parameters accurately and efficiently to achieve close-to-optimal end-to-end data transfer throughput with very low overhead. \algoName~uses historical data to derive network specific models of transfer throughput based on protocol parameters. Then by running sample transfers, we capture current load on the network which is fed into these models to increase the accuracy of our predictive modeling. Our experimental analysis over a variety of network settings shows that HARP outperforms existing solutions by up to 50\% in terms of the achieved throughput.

\ifCLASSOPTIONcaptionsoff
  \newpage
\fi



%
\bibliographystyle{IEEEtran}
\footnotesize
\bibliography{references}

%

\vspace{-1.67cm}



\end{document}